\documentclass[fleqn,usenatbib]{mnras}

\usepackage{newtxtext,newtxmath}

\usepackage[T1]{fontenc}

\DeclareRobustCommand{\VAN}[3]{#2}
\let\VANthebibliography\thebibliography
\def\thebibliography{\DeclareRobustCommand{\VAN}[3]{##3}\VANthebibliography}

\usepackage{graphicx}	
\usepackage{amsmath}	
\usepackage{booktabs}
\usepackage{multirow}
\usepackage{color,soul}

\newcommand{\simspin}[1]{\textsc{SimSpin}#1} 

\newcommand{\eagle}[1]{\textsc{Eagle}#1} 
\newcommand{\sami}[1]{\textsc{SAMI}#1} 
\newcommand{\msol}[1]{M$_{\odot}$}

\title[From Particles to Pixels]{From Particles to Pixels: How many particles do I really need to construct stellar kinematic mock observational measurements?}

\author[K. E. Harborne et al.]{
K. E. Harborne,$^{1,2}$\thanks{E-mail: katherine.harborne@icrar.org}
C. del P. Lagos,$^{1,2}$
S. M. Croom,$^{2,3}$
J. van de Sande, $^{2,4}$ 
A. Ludlow, $^{1,2}$
\newauthor{
 R. S. Remus,$^{5}$
 L. C. Kimmig, $^{5}$
 and C. Power$^{1,2}$
} \\
$^{1}$International Centre for Radio Astronomy (ICRAR), M468, The University of Western Australia, 35 Stirling Highway, \\ 
Crawley, WA 6009, Australia\\
$^{2}$ARC Centre of Excellence for All Sky Astrophysics in 3 Dimensions (ASTRO 3D) \\
$^{3}$Sydney Institute for Astronomy (SIfA), School of Physics, A28, The University of Sydney, NSW, 2006, Australia \\
$^{4}$School of Physics, University of New South Wales, Kensington, NSW, 2032, Australia \\
$^{5}$Universit\"ats-Sternwarte, Fakult\"at f\"ur  Physik, Ludwig-Maximilians Universität, Scheinerstr. 1, 81679 M\"unchen, Germany \\
}

\date{Accepted XXX. Received YYY; in original form ZZZ}

\pubyear{2015}

\begin{document}
\label{firstpage}
\pagerange{\pageref{firstpage}--\pageref{lastpage}}
\maketitle

\begin{abstract}
This work considers the impact of resolution in the construction of mock observations of simulated galaxies. 
In particular, when building mock integral field spectroscopic observations from galaxy formation models in cosmological simulations, we investigate the possible systematics that may arise given the assumption that all galaxies above some stellar mass limit will provide unbiased and meaningful observable stellar kinematics. 
We build a catalogue of N-body simulations to sample the range of stellar particle resolutions within the \eagle{} \texttt{Ref0050N0752} simulation box and examine how their observable kinematics vary relative to a higher-resolution $N$-body control. 
We use these models to compile a table of the minimum number of particles-per-pixel to reach a given uncertainty in the fitted line-of-sight velocity distribution parameters. 
Further, we introduce a Voronoi-binning module to the mock observation code, \simspin{,} in order to meet these minimum numbers.  
Using \eagle{}, we show the impact of this shot noise on the observed spin-ellipticity plane and the recovery of this space when observations are binned with increasing numbers of particles. 
In conclusion, we advise binning mock images to meet at least 200 particles-per-pixel to avoid systematically under-estimating the velocity dispersion along a given line-of-sight. 
We demonstrate that this is important for comparing galaxies extracted from the same simulation, as well as between simulations of varying mass resolution and observations of real galaxies.
\end{abstract}

\begin{keywords}
galaxies: kinematics and dynamics -- methods: numerical -- galaxies: evolution
\end{keywords}



\section{Introduction}

When designing a method to test an astrophysical question, it is prudent to first consider the required “resolution” of our data. 
In the scenario of measuring stellar kinematics observationally, the spatial and spectral resolution of the observing instrument must be assessed.
A necessary signal-to-noise ratio and minimum acceptable level of seeing are prescribed to achieve some level of precision in the recovered kinematics. 
Similarly, for simulations of galaxy formation and evolution, the mass resolution and gravitational softening of a model must be considered, alongside the relative sampling of the underlying dark matter, to understand within what bounds the kinematics measured can be trusted above numerical effects and the limitations of sub-grid prescriptions of feedback and cooling. 

For astronomers working at the intersection between observation and theory, the consideration of resolution – in particular, at what resolution our observational and modelled stellar kinematic results are comparable in "mock observation" form – is a yet unanswered question. 
In this work, we hope to begin the process of assessing required resolution parameters to produce unbiased mock observations.  

Extensive literature on the impact of resolution within integral field spectroscopic (IFS) instrumentation can be found, particularly for the measurement of stellar kinematics. 
Primarily, unbiased spectral fitting methods to recover stellar kinematics were championed by \cite{Kuijken1993Adecomposition}, who demonstrated that a minimum signal-to-noise is required. 
This was built upon further by \cite{Cappellari2003Vorbin}, who introduced the methodology of Voronoi binning, using adaptively sized bins based on a minimum signal-to-noise constraint to acquire unbiased spectral fits to the line-of-sight velocity distribution (LOSVD). 
As a result, users could be confident that the resulting uncertainty across a 2D spectroscopic map were not systematically biased by low S/N pixels across the image.  

Beyond the signal-to-noise, other effects such as the variation in aperture size from observation to observation in a survey suite have been considered and corrected \citep[e.g.][]{Jorgensen1995Spectroscopyclusters, vandeSande2013stellargalaxies, FalconBarroso2017Stellarcorrections, vandeSande2017SAMIsurvey}, as well as the effects of seeing conditions \citep[e.g.][]{dEugenio2013Fast183, Graham2018SDSSproperties, Greene2018MaNGAAMGalaxies, Harborne2020Recoveringkinematics, vanHoudt2021Legac, Derkenne2024MAGPISRs, MunozLopez2024MUSESeeing}. These effects are particularly important for IFS surveys where many galaxies of various sizes are observed at a variety of different seeing conditions with the eventual aim of being compared within the same parameter space, e.g. SAMI \citep[the Sydney-AAO Multi-object Integral Field Spectrograph;][]{Croom2012SAMIOverview}, MaNGA \citep[Mapping Nearby Galaxies at Apache Point;][]{Bundy2015MaNGA}, CALIFA \citep[the Calar Alto Legacy Integral Fields Area survey;][]{Sanchez2012CALIFAOverview}, MAGPI \citep[Middle-Ages Galaxy Properties using Integral field spectroscopy;][]{Foster2021MAGPIOverview}, LEGA-C \citep[Large Early Galaxy Astrophysics Census;][]{vanderWel2016LEGAC, Bezanson2018LEGA-C} .
Such corrections ensure that galaxy observations are comparable from night to night, as well as to one another between surveys.

Similarly widespread is the literature considering the effects of resolution on simulations of galaxy formation and evolution. 
It should be clear that a simulation -- wherein point mass particles trace the distributions of gas and stars in tens of thousands of galaxies -- will be unable to reproduce physics on the scale of individual stars when the mass of a single particle is of order $\sim 10^4 - 10^6$ \msol{} (which are typical values for contemporary simulations). 
Further, the equations of gravity must be softened on small scales, usually of order 100s of pc. 
This is done to prevent large angle deflections during two-body scattering so that low-order integration schemes can be used. 
The kinematics and spatial distribution of particles below the softening length (or several multiples of it) will not be simulated reliably. 
Such numerical effects are well known and have been comprehensively studied in the context of dark matter haloes. 
\citeauthor{Power2003innerstudy} (\citeyear{Power2003innerstudy}; see also \citealt{Ludlow2019NumericalHaloes}), for example, derived local conditions for convergence\footnote{Convergence is typically defined as the (radial or mass) scale above which the properties of a simulated object (i.e., a DM halo or galaxy) are consistent with the same object simulated at a higher resolution.} in the circular velocity profiles of haloes.


Convergence in the properties of simulated galaxies is more difficult to establish due to the complex coupling of numerical and subgrid parameters. 
However, systematic tests were carried out for the \eagle{} galaxy formation model in \citeauthor{Ludlow2020NumericalHaloes} (\citeyear{Ludlow2020NumericalHaloes}; for the stellar mass function, galaxy sizes and rotation curves, among other properties), and in \citeauthor{Lagos2017AngularEAGLE} (\citeyear{Lagos2017AngularEAGLE, Lagos2018Thegalaxies}; for the baryonic and neutral gas angular momentum and spin-parameter of galaxies). \citeauthor{Ludlow2019EnergySizes} (\citeyear{Ludlow2019EnergySizes}; and further in \citealt{Ludlow2021Spuriousparticles, Wilkinson2023impactdiscs}) showed that the number of dark matter halo particles must also be considered, as too few particles ($\lesssim 10^6$) result in spurious collisional heating of galaxies, which affects (among other properties) their sizes, morphologies, stellar velocity dispersion profiles, and angular momenta, ultimately modifying the normalization and slope of fundamental galaxy scaling relations \citep[e.g.][]{Ludlow2023Spuriousformation}.


When bringing together our simulations and observations through the construction of mock observables, is it sufficient to take galaxies classed as ‘converged’ within the traditional theory framework? From the observational point of view, we can manufacture our mock observations to meet some maximum added noise requirement in mimicking the effects of signal-to-noise; we can control for aperture effects and seeing conditions across the sample. Is it sufficient to ensure that observational pixels are above the size of the gravitational softening employed within the simulation?  

We have seen an explosion in popularity of direct comparisons between observations and simulations in recent years \citep[e.g.][to name but a few of the methods of constructing mock integral field spectroscopic data]{vandeSande2019SAMISimulations, Harborne2020SimSpin, WaloMartin2020KinematicAccretion, Bottrell2022RealisticIFS, Nanni2023iMaNGASSP, Sarmiento2023MaNGIAanalysis, Harborne2023SimSpin}. 
Such works enable us to evaluate the physical drivers of kinematic transformation.
Feedback models from active galactic nuclei (AGN) can be seen to influence the spin of massive galaxies in a manner beyond what we observe \citep{vandeSande2019SAMISimulations, Lagos2018Thegalaxies}. We have also seen that environment and galaxy interactions are important in shaping kinematic morphology in some models, \citep{Valenzuela2024ShapesMagneticum, Lagos2021Diversesimulations, Foster2021MAGPIOverview, Schulze2018KinematicsMagneticum}, but less important in others \citep{Nanni2024iMANGA3}. 
Such results prompt observational programs to chase down an answer to these comparisons by extending our resolved view of stellar kinematics back to higher redshift to explore the balance between these processes \citep[e.g.][]{Foster2021MAGPIOverview}.
Similarly, for theorists modifying simulation design to explore how better to capture stellar and AGN feedback effects in the interstellar medium,\citep[e.g.][]{Ploeckinger2024ResolutionISM, Nobels2024SubgridSF, Chaikin2023SubgridAGN} such comparisons will be imperative for testing success.

Not limited to stellar kinematics, mock observations have been an essential tool for bridging observations and theory.
For example, the diversity of dwarf galaxy rotation curves using atomic hydrogen \citep[e.g.][]{Oh2011DMTHINGS, Adams2014DwarfKinematics, Oman2015DiversityProblem, Oman2019DiversityProblem} may imply we need a new cosmological model \citep{SantosSantos2020BaryonRotationCurves, Roper2023RCDiversityCuspCore}, or that the observational biases of HI observations lead naturally to a variation in observed rotation curves \citep{Downing2023RotationCurves, Sands2024DMProfiles}. 
Without mock observations, it is impossible to distinguish between these possibilities. 
Similarly, in exploring the impact of mergers and interactions on galaxy evolution, mock observations confirm the importance of projection effects on the observational interpretations of halo mass \citep[e.g.][]{ContrerasSantos2022GalaxyPairs300, ContrerasSantos2023GalaxyPairs300_2}.
The impact of dust obscuration, spatial resolution and surface-brightness limits on our understanding of galaxy merger rates across cosmic time are all enhanced by the construction of mock observations \citep[e.g.][]{Marini2024eROSITAGroupPred, Bottrell2019DeepRealism, Rodriguez-Gomez2019TNGobservations, BaesCamps2015SKIRT}.
While this is an incomplete list of examples, clearly mock observations are an important tool for understanding many areas of galaxy evolution, and as such it is important that we understand how biases may propagate in their construction. Here, we will focus specifically on the recovered stellar kinematics through the construction of mock IFS observations.

In the construction of such mocks, a single galaxy's particles are gridded into pixels comparable to the resolution of the telescope being mimicked. 
Other observational limitations, such as projection and the effects of seeing conditions, can further be added to increase the realism of such data and the resulting mock observations can be analysed using the same methods as is used for real, observational data. 
Measured properties are then considered to be a more faithful comparison to observed values, with systematic differences now minimised. 

When constructing a projected pixel grid into which the properties of the particles contained are summarised to match a given telescope resolution, how many particles are necessary to determine an unbiased fit to these line-of-sight velocity distributions? 
If we stay fixed to a given pixel size with a model projected at a fixed distance, will a galaxy of mass $\approx 5 \times 10^9$ \msol{} have equal uncertainty in the computed the line-of-sight velocity distribution (LOSVD) to a more massive companion from the same simulation, for example?
This particles-per-pixel value will be dependent on distance to which a given system is projected, its underlying morphology and projection angle and the resolution of the mock instrument used to observe the galaxy (i.e. the fixed angular pixel size). 
Is it fair to compare similar objects from different simulations with different particle resolutions if their "mock" telescope is consistent, or do further elements need to be controlled for?

In this work, we present a convergence test designed to explore the effects of numerical resolution on the construction of mock observations in systems that would be classically numerically converged. 
We examine how well we can recover the LOSVD in such systems and explore the particles-per-pixel value necessary to overcome the effects of shot-noise in the recovery of the LOSVD. 
We go on further to explore methods of mitigating these statistical uncertainties through the use of Voronoi-binning pixels to some ``particle-to-pixel" limit.
We present the reader with a table of suggested particles-per-pixel limits, depending on the kinematic parameters you wish to recover. 
The aim of this work is to provide a resource to astronomers wishing to build observational measurements of simulated galaxies in consistent ways - both with observations and other simulations. 
While the current generation of galaxy formation models are soon to be superseded, it is important that we can compare between the models, older-generations to new, to understand the relative improvements made in sub-grid implementations.
This will be impossible without accounting for the possible biases introduced by the "mock" observation generation process. 

In \S \ref{sec:methods}, we outline the models built for this study, and in \S \ref{sec:convergence} we detail the methods used to quantify the level of statistical uncertainty in a variety of properties measured from these models. 
In \S \ref{sec:implications}, we discuss reasonable methods for minimising the effects of shot noise on the mock-observed properties, presenting a new \simspin{} method of Voronoi-binning pixels in a similar way to observers in an attempt to maximise signal-to-noise.
We further discuss the implications this procedure may have on previous evaluations of the slow rotator fractions presented for galaxy formation simulations in the literature.
In \S \ref{sec:conclusions}, we present our conclusions including a tabulated summary of the minimum number of particles necessary to recover observationally comparable properties within reasonable statistical uncertainty.


\section{Method}
\label{sec:methods}

\subsection{The models}
We begin by building a number of two-component simulated systems for which the kinematics of the model are well sampled, which we call our \texttt{control} systems.  
$N$-body models of stellar discs within a DM halo and elliptical systems within a DM halo have been constructed using the initial conditions code \texttt{GalIC} \citep{Yurin2014AnEquilibrium} and evolved in a static analytic potential using a modified version of \texttt{Gadget2} \citep{Springel2005TheGADGET-2}. 
Each \texttt{control} galaxy structure is sampled by $6.5 \times 10^{6}$ ``stellar'' particles of mass $1 \times 10^{4}$ \msol{}, such that the total stellar mass of the system is $M_{*} = 6.5 \times 10^{10}$ \msol{} totally contained in either a dispersion dominated bulge or rotation supported disc.
This stellar structure is embedded within a dark matter halo of mass $\approx 2.5 \times 10^{12}$ \msol{}, such that the system sits at the peak galaxy-formation efficiency of the halo mass function in \cite{Behroozi2013AverageHalos}. 

We further create a number of systems at the intermediate resolution of the \eagle{} galaxy formation model. 
Due to the range of masses with which a star may be born from a given gas particle, a galaxy of stellar mass $M_* = 6.5 \times 10^{10}$ \msol{} within the intermediate-resolution {\textsc EAGLE} simulation runs can be sampled by $\sim200,000$ to $\sim260,000$ stellar particles. 
Five models are made at the minimum, the 16th, 50th, 84th percentiles, and the maximum possible number of particles for a galaxy of the given mass (named \texttt{p0}, \texttt{p16}, \texttt{p50}, \texttt{p84}, \texttt{p100} respectfully) for each of the two extremes of rotation support (``bulge'' and ``disc''). 
These systems have an identical stellar mass, velocity structure and density distribution to their respective \texttt{control} system, but are sampled using a lower number of higher mass stellar particles. 
The particles in the lowest resolution models have a mass of $3.34 \times 10^{5}$ \msol{}, decreasing to a mass of $2.48 \times 10^{5}$ \msol{} in the highest resolution model. 
The specifics of each model flavour are described in Table \ref{tab:Nbody_models}.
Each system sits in the same DM mass halo as the original control model and is initialised with an equal halo-to-stellar particle mass fraction of 5.36, i.e. $m_{\text{DM}} = 5.6 \times m_{*}$, where $m$ denotes a particle's mass. This live potential is replaced with an analytic DM halo of this mass before the system is evolved.
This suite of idealised simulations forms the basis of our convergence test.

\begin{table}
    \centering
    \caption{Parameter description of $N$-body models built using \texttt{GalIC} and \texttt{Gadget2}. For each row, two models have been made, one in the Hernqiust bulge density distribution and the other with a sech$^2$ radial disc density distribution.}
    \label{tab:Nbody_models}
    \begin{tabular}{lcrr}
       \hline \textbf{Name}  & \textbf{Type} & \textbf{No. Particles}  & \textbf{Mass per particle, M$_{\odot}$} \\ \hline
       \texttt{control} & bulge / disc & 6,500,000 & $1 \times 10^{4}$\\
       \texttt{p100} & bulge / disc & 261,993 & $2.48 \times 10^{5}$\\
       \texttt{p84} & bulge / disc & 244,800 & $2.66 \times 10^{5}$\\
       \texttt{p50} & bulge / disc & 225,270 & $2.89 \times 10^{5}$\\
       \texttt{p16} & bulge / disc & 206,369 & $3.15 \times 10^{5}$\\
       \texttt{p0} & bulge / disc & 194,358 & $3.34 \times 10^{5}$\\ \hline
    \end{tabular}
\end{table}

The DM halo and bulge are initialised as spherically symmetric, isotropic models with a Hernquist density profile \citep{Hernquist1990} described by:
\begin{equation}
    \rho(r) = \frac{M}{2 \pi} \frac{a}{r(r+a)^3}, \label{eq:hernq}
\end{equation}
where $a$ is the scale factor\footnote{GaLIC parameterises the scale factor, $a$, such that the inner regions resemble a Navarro-Frenk-White profile \citep{Navarro1996}, with $c$=10 and $M_{200} = 2.5 \times 10^{12}$\msol{}.} 
Here, $M$ is the total mass of the component (i.e. $M_{*}$ for the bulge and $M_{DM}$ for the halo).
The bulge scale length defined as a fraction of the halo scale length such that the $a_{\text{halo}} = 37.9$ kpc and $a_{\text{bulge}} = 0.2 \times a_{\text{halo}} = 7.6$ kpc. 

The disc is modelled with an axisymmetric velocity profile, exponential radial density profile and sech$^{2}$-profile in the vertical direction, described by:
\begin{equation}
    \rho(R,z) = \frac{M_*}{4 \pi z_{d} R_{d}^2}\text{sech}^2\left( \frac{z}{z_{d}} \right)\text{exp}\left( - \frac{R}{R_{d}} \right).
\end{equation}
where the disc is aligned such that the $z$-axis sits perpendicular to the plane, $R$ is the radius from the $z$-axis and $z$ is the height off that plane. 
Again, $M_*$ is the total stellar mass of the disc, while $R_{d}$ is the disc scale length and z$_{d}$ is a radially constant disc scale height. 
For the discs built for this experiment, we have $z_{d} = 0.8$ kpc and $R_{d} = 4.2$ kpc, such that $z_{d}/R_{d}$ = 0.2 in line with expectations for typical discs. 
These discs are embedded within the DM halo as described by the Hernquist density profile in Eq. \ref{eq:hernq}.

Using the optimisation procedure outlined in \cite{Yurin2014AnEquilibrium}, the velocities of each particle in the system is initialised to satisfy the collisionless Boltzmann equation. 
To avoid the need to evolve the system within a computationally expensive halo of dark matter particles that may cause an exchange in energy between the collisionless distributions \citep{Ludlow2021Spuriousparticles}, these systems are evolved in an analytic potential for 10 Gyr using \texttt{Gadget2}. 

\begin{figure}
    \centering
    \includegraphics[width=\linewidth]{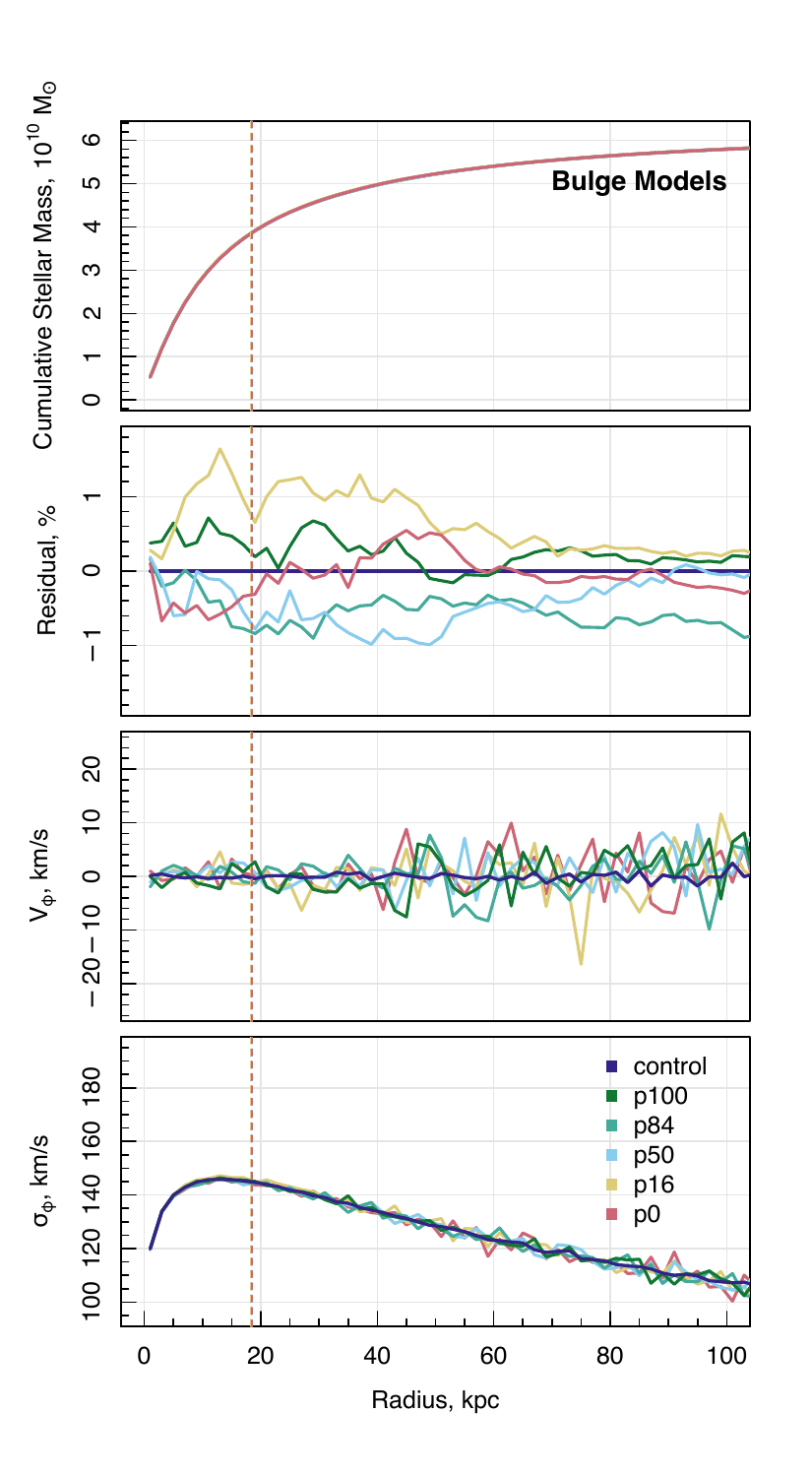}
    \caption{Radial analyses of the six different systems in the bulge suite of models, from the \texttt{control} model through to the lowest resolution \texttt{p0}. We compute each property within cylindrical bins of radius, with the cylinder aligned with the angular momentum vector. Bins of width 2 kpc have been used from 0-100 kpc. The dashed vertical line denotes the physical half-mass radius for each model. (\textit{Row 1}) The cumulative mass distribution in units of $10^{10}$ \msol{}. (\textit{Row 2}) The residual differences between the models of each resolution is within 1\% of the \texttt{control} model across the full radial range. (\textit{Row 3}) The mean azimuthal velocity in each radial bin measured in units of km/s. (\textit{Row 4}) The azimuthal velocity dispersion in each radial bin in units of km/s.}
    \label{fig:bulge_cumulative_mass}
\end{figure}

\begin{figure}
    \centering
    \includegraphics[width=\linewidth]{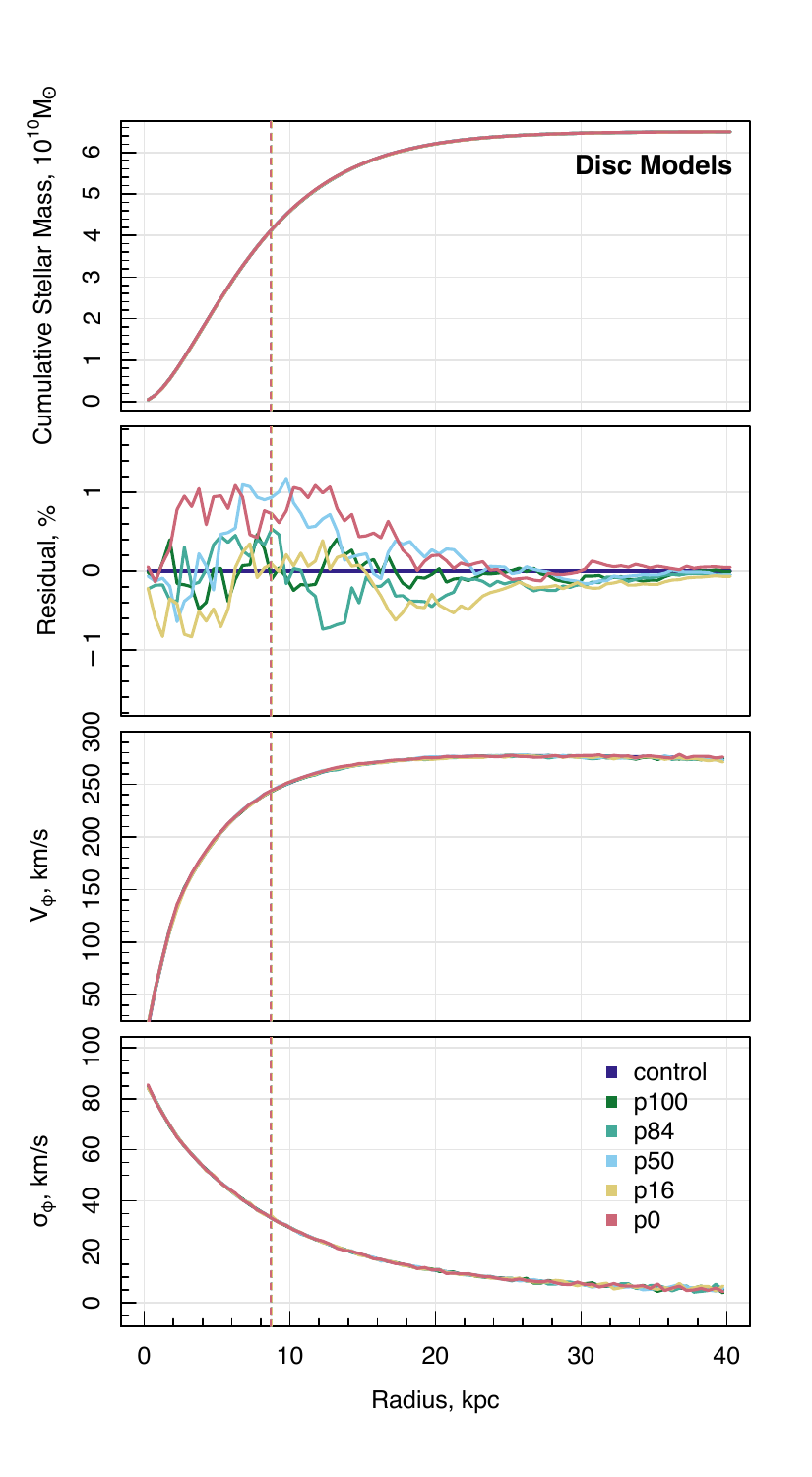}
    \caption{As in Fig \ref{fig:bulge_cumulative_mass}, but for six different systems in the disc suite of models. Cylindrical bins of radius of width 0.5 kpc have been used from 0-40 kpc, where the disc ends. The dashed vertical line denotes the physical half-mass radius for each model. (\textit{Row 1}) The cumulative mass distribution in units of $10^{10}$ \msol{}. (\textit{Row 2}) Demonstrating that the residual differences between the models of each resolution is within 1\% of the \texttt{control} model across the full radial range considered. (\textit{Row 3}) The mean azimuthal velocity in each radial bin measured in units of km/s. (\textit{Row 4}) The azimuthal velocity dispersion in each radial bin in units of km/s.}
    \label{fig:disk_cumulative_mass}
\end{figure}

We measure the cumulative mass distribution in radial bins for the simulations at each resolution for the bulge models in Figure \ref{fig:bulge_cumulative_mass} and for the disc models in Figure \ref{fig:disk_cumulative_mass}. 
Properties are measured in bins of width 2 kpc and 0.5 kpc in the bulge and disc models respectfully, with each bin containing more than 500 particles. 
The azimuthal ``rotational'' velocity ($V_{\varphi}$) and dispersion ($\sigma_{\varphi}$) measured in cylindrical shells of radius, co-axial with the angular momentum vector of the system, are also shown in the panels of each figure.
These distributions demonstrate that each model is sufficiently well-resolved within the half-mass radius, R$_e$, shown by the dashed vertical line for each model. 
In the case of the disc model, we see very little variation between the systems, with differences in the cumulative mass of order $\sim 1\%$. 
For the bulge, where dispersion dominates the kinematics, we see a greater spread in velocity profiles between each model, though well-scattered about the true control model results. 
This justifies the need to investigate these two extremes of the kinematic parameter space.
Most systems will sit in-between these two idealised modes. 

This suite of models provides the basis of our convergence test.
However, it is also important to understand the impact of this test on the broader ecosystem of hydrodynamical, galaxy formation models. 
For this reason, we will also be examining galaxies extracted from the \eagle{} simulation in \S \ref{sec:implications}.
Galaxies of various morphologies above a stellar mass limit of $10^{10}$ \msol{} have been extracted from the \texttt{RefL0050N0752} run. 
This is a 50 Mpc co-moving box of Universe run with the reference \eagle{} model of galaxy formation and evolution. 
We refer the reader to the works of \cite{Schaye2015EAGLEenvironments} and \cite{Crain2015EAGLE} for the specific descriptions of this simulation and here highlight some specifics regarding the resolution. 
Nominally, gas particles have a mass of $1.8 \times 10^{6}$ \msol, with particles tracing the dark matter distribution described by collisionless particles of mass $9.7 \times 10^{6}$ \msol. 
Stars are born from ``cold'', dense gas when the critical, metallicity-dependent density threshold is satisfied. 
The mass of these stars can vary and is dependent on the density of the gas at the time and space of star formation. 
In essence, this means that individual stellar particles may have masses from $10^{5}$ \msol{} to $10^{9}$ \msol, and a galaxy of a given mass will be made of a variable number of stellar particles. 
Using the $N$-body models above, we choose to explore the lower end of this mass range limit, investigating what would be considered as the highest resolution models in the simulation.

Using the public \eagle{} database \citep{McAlpine2016EAGLEdatabase}, we have extracted 50 kpc spheres around the centre of potential of identified subhaloes from the $z = 0$ snapshot of the \texttt{RefL0050N0752} box. 
Only particles associated with a given subhalo have been used for further analysis. 
These systems meet a minimum stellar mass cut of $1 \times 10^{10}$ \msol{} within the subhalo, following the work of \cite{Lagos2021Diversesimulations}. 
Objects where the half-mass radius falls outside the 50kpc sphere are removed from analysis (1/481), as are a few systems with insufficient numbers of particles to be Voronoi binned (6/481; of relevance to \S \ref{sec:implications}). 
This leaves us with 474 subhaloes from the \eagle{} simulation for further investigation into the impact of particle-per-pixel limits on the recovery of observable kinematics. 
We note that systems between M$_{*}$ $\sim 10^{10}$-$10^{11}$\msol{} may be subject to spurious heating within R$_{e}$ due to interactions between stellar and dark matter particles \citep{Ludlow2023Spuriousformation}. 
This does not significantly change our results on the topic of particles-per-pixel, as will be seen in \S \ref{sec:implications} and further in appendix \ref{app:highres-dm}, where we re-measure the observable kinematics of galaxies extracted from the higher-resolution dark matter re-simulation of \eagle{} box RefL0050N0752 from \cite{Ludlow2023Spuriousformation}.

\subsection{Mock observations}

\begin{figure*}
    \centering
    \includegraphics[width=0.9\linewidth]{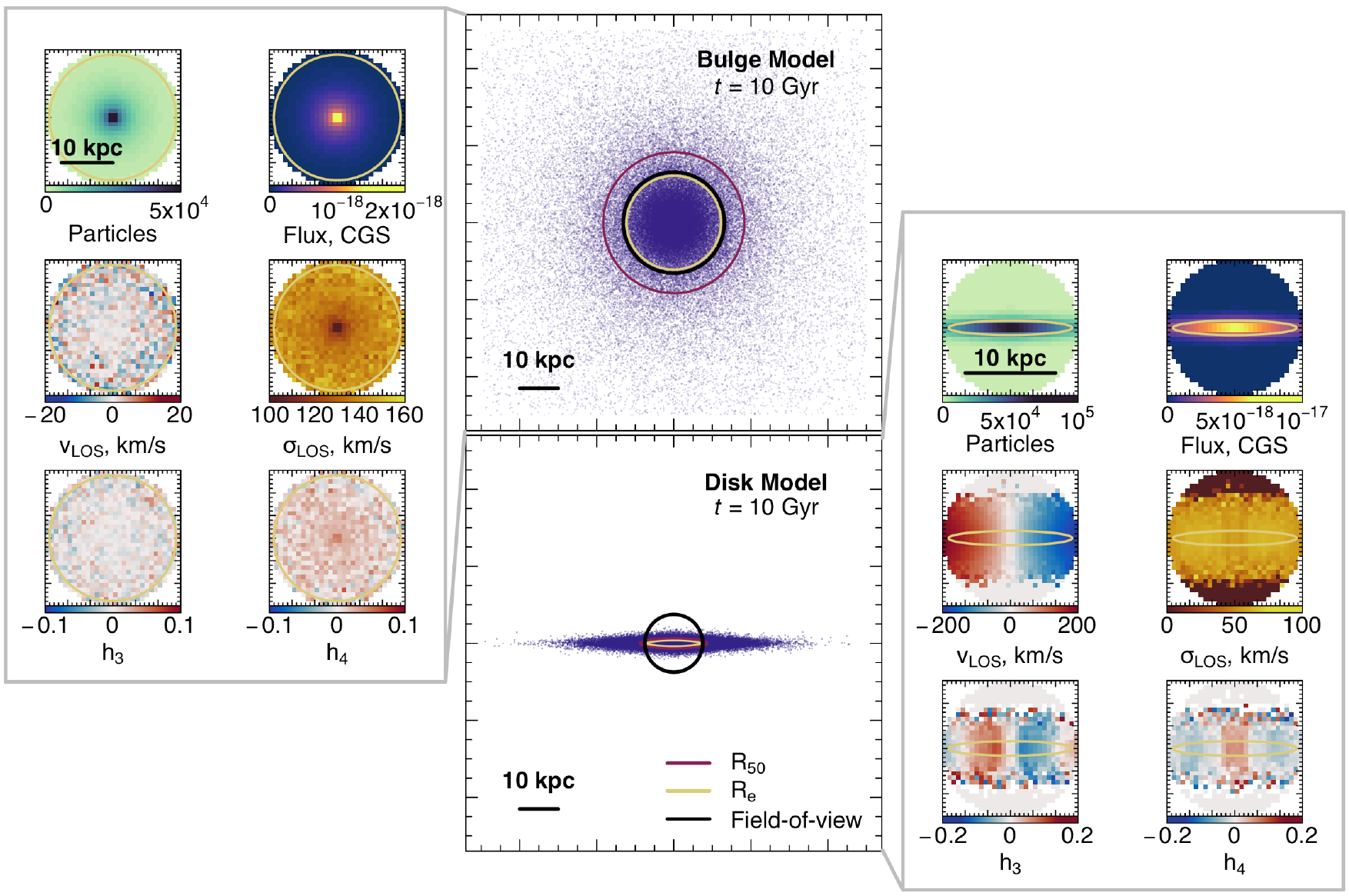}
    \caption{Demonstrating mock observations of the $N$-body models used in the convergence study. (\textit{Centre}) Above, we show a map of the particle distribution of the bulge model. Below, the same for the disc model. In each case, we mark the measured half-mass radius (R$_{50}$) in red, half-light radius (R$_e$) in yellow and the SAMI field-of-view region in black. (\textit{Left}) Example particle maps, $r$-band flux and kinematic maps produced by \simspin{} for the bulge model. (\textit{Right}) The same demonstration of \simspin{} output maps, but for the disc model.}
    \label{fig:mock_observations}
\end{figure*}

We generate mock integral field spectroscopic (IFS) observations using \simspin{} \citep{Harborne2023SimSpin}.
This is an open-source, well-documented code\footnote{\url{https://kateharborne.github.io/SimSpin/}} that has been developed to generate comparable data products for a range of different input simulation types and output telescope configurations. 
We direct readers to the summary paper for an overview of this code, but outline the main methodology here. 

Input simulations of galaxies, as sampled by particles or cells, have a number of tagged properties that are used by \simspin{} including positions ($x, y, z$), velocities ($v_x, v_y, v_z$), and masses. 
For hydro-dynamical inputs, stellar particles will also have metallicity and age information which can be used to attach a spectral template to that particle.
For N-body models, \simspin{} assigns a single value of stellar age and metallicity to individual velocity components (i.e. bulge and disc) such that a spectral template can be attached to these particles also. 

\simspin{} grids the particles into bins according to (1) the requested telescope configuration (i.e. the spaxel size, the wavelength resolution, the field-of-view, etc.) and (2) observing strategy (i.e. the projected distance to the centre of the galaxy, the projected inclination angle, the atmospheric conditions, etc.)
The LOSVD of particles in each pixel bin position is fit, weighted by the luminosity of each particle as computed from their assigned spectral template and as observed at the conditions described by (1) and (2).  
The mock observation may then be convolved to match the requested atmospheric conditions and is returned in FITS format for further processing. 

In this work, we make use of \simspin{} to explore the effects that the number of particles-per-pixel has on the recovery of the observable kinematics. 
Throughout this work, we will be using a mock telescope that matches the configuration of SAMI \citep{Croom2012SAMIOverview}. 
Each bundle of multimode fibres on this instrument subtend a diameter of 15'' on the sky, which dictates the field-of-view of our mock observations. 
Most of the absorption lines used for kinematic measurements fall within the range of the blue arm of the AAOmega dual beam spectrograph and so these specifications are used for the mocks. 
A 580V grating is mounted on the blue arm of AAOmega, which gives a resolution of R $\sim 1700$ and wavelength coverage of 3700 - 5700 \AA{} \citep{Sharp2006AAOmega}. 
The associated line-spread function is well-approximated by a Gaussian with full-width half-maximum of 2.65 \AA{} \citep{vandeSande2017SAMIKinematics}. 
The cubes constructed have a spatial pixel scale of 0.5'' \citep{Sharp2015SAMIholes, Green2018SAMIDR1}.
\simspin{} has the configuration for the SAMI instrument as a predefined type and this is used for all IFS mocks throughout this work.

We have measured the half-light radius, R$_e$, of each system by constructing a KIDS-like $r$-band image containing the entirety of each model (i.e. the size of each image is variable in size, but with spatial pixel resolution fixed at 0.2'' and ideal seeing conditions with no sky noise added).
Using the code ProFound \citep{Robotham2018ProFounddata}, we construct isophotal ellipses for each mock image and fit for R$_e$, as would be done in real observations. 

For the $N$-body simulations that form our convergence test suite, we build our mock observations such that the half-light radius (R$_e$) falls just within the field-of-view of SAMI. 
We use the photometric information from ProFound to determine the necessary distance to project our idealised $N$-body systems such that they meet this requirement. 
For the bulge model, this distance is $z$ = 0.09 (430 Mpc or 1.75 kpc/''); for the disc model, this distance is $z$ = 0.05 (227 Mpc or 1 kpc/'').
Both objects are inclined edge-on at 90 degrees. 
The same distance is used for all models within the suite of resolutions, as all have the same morphological structural parameters, as is evident in Figures \ref{fig:bulge_cumulative_mass} and \ref{fig:disk_cumulative_mass}.
Examples of the resulting mock images can be seen in Figure \ref{fig:mock_observations}.

We use the resulting kinematic maps to explore how well we can recover the first four moments of the LOSVD with varying numbers of particles-per-pixel. 
In this work, we parameterise the LOSVD via the method outlined in \cite{Marel1993GaussHermite}.
The distribution is assumed to be Gaussian with higher order corrections given by:
\begin{equation}
    f(\nu_{los}) \propto e^{-\frac{1}{2} \omega^2} [1 + h_3 H_3(\omega) + h_4 H_4 (\omega)],
    \label{eq:LOSVD}
\end{equation}
where,
\begin{align}
\omega &= (\nu_{los} - \bar{V})/\sigma, \label{eq:omega} \\
H_3(\omega) &= \frac{1}{\sqrt{6}} \left( 2\sqrt{2} \omega^3 - 3\sqrt{2} \omega \right), \\
H_4(\omega) &= \frac{1}{\sqrt{24}} \left( 4 \omega^4 - 12 \omega^2 + 3 \right). 
\label{eq:h4}
\end{align}

The function, $f$ describes the LOSVD, where $\nu_{los}$ is the velocity along the line of sight, with $\bar{V}$, $\sigma$, $H_3$ and $H_4$ representing the moments of this distribution which can be approximated as Gaussian when $H_3$ and $H_4$ tend to zero.
These latter terms are the third and fourth Hermite polynomials, for which the coefficients $h_3$ and $h_4$ are often used within the literature to describe the higher-order terms of the LOSVD.
We expect that, to some degree at which shot-noise is dominant, the recovered fit for these moments will be dependent on the number of particles-per-pixel in a given spaxel and the required number of particles will be greater for accurate recovery of the higher-order terms.
We explore this further in the convergence test presented in Section \ref{sec:convergence}.

It is noted that, when running kinematic fitting codes such as pPXF \citep{Cappellari2004pPXF, Cappellari2017pPXF, Cappellari2023pPXF}, a Gaussian function can also be used to more accurately recover the first two moments of the LOSVD (e.g. in pPXF v9.4.1, the keyword `\texttt{moments}` can be set at `2` or `4`)\footnote{\url{https://pypi.org/project/ppxf/}}.
It is common to run a kinematic fit using both functions (Gaussian and a Gauss-Hermite polynomial) in order to compute summary properties of observational kinematics within R$_{e}$ using the two-moment fit.
\simspin{} enables the user to select between a two-moment and four-moment fit for the LOSVD when building mock observations, in keeping with the methodology of the SAMI survey.
As in \ref{eq:LOSVD}, for a Gaussian fit we just consider the first term of the function:

\begin{equation}
    f(\nu_{los}) \propto e^{-\frac{1}{2} \omega^2},
    \label{eq:Gaussian}
\end{equation}

where $\omega$ is defined as in Eq \ref{eq:omega}. 
For all fits throughout this work, we consider the effect of the number of particles-per-pixel on both the Gauss-Hermite (Eq. \ref{eq:LOSVD}) and Gaussian (Eq. \ref{eq:Gaussian}) fits to the LOSVD. 

A similar methodology is used for the galaxies extracted from the \eagle{} simulation.
Each galaxy is projected at a fixed 70 degree inclination, such that the effects of inclination are kept independent of mock resolution.
For the \eagle{} systems, we vary the distance to the object until the half-light radius falls within the central 28 pixels of the image.  
Kinematic maps, such as those shown in Figure \ref{fig:mock_observations}, are produced for each observation through the same fit of Equation \ref{eq:LOSVD}.
From these maps, rather than just considering the moments of the LOSVD, we compute a number of observable kinematic parameters that are commonly used in the literature to classify the kinematic morphology of galaxies. 
As such, we define a number of these parameters below, including $\lambda_R$, $V/\sigma$ and $j_*$, within 1 R$_e$.

The observed stellar spin parameter, $\lambda_R$, is a proxy that quantifies the observed projected stellar angular momentum \citep{Emsellem2007Saurongalaxies} which is computed via,
\begin{equation}    
    \lambda_R = \frac{\sum_{i=1}^{n_b} F_i R_i |V_i|}{\sum_{i=1}^{n_b} F_i R_i \sqrt{V_i^2 + \sigma_i^2}},
    \label{eq:lr}
\end{equation}
where $F_i$ is the observed ($r$-band) flux, $R_i$ is the elliptical radius of the bin centre relative to the centre of the object, $V_i$ is the line-of-sight (LOS) velocity and $\sigma_i$ is the LOS velocity dispersion.
These $V_i$ and $\sigma_i$ are taken from the two-moment Gaussian fits to the LOSVD.
The sum of different properties is made across $n_b$ bins whose centres fall within the effective radius, R$_e$ of the object (whether those bins be Voronoi bins or individual spaxels). 
In particular, the $\lambda_R$ parameter is used to categorise galaxies into different kinematic morphology classifications, in particular ``slow'' and ``fast'' rotators. 
Throughout this work, we will be using the slow rotator (SR) fraction, or the relative number of SR's to the total number of systems, to demonstrate the impact of particle-per-pixel values. 
As such, it is important to define the SR classification, for which we use the convention popularised by \cite{Cappellari2016AnnualReview}.
We note that for a holistic view of the kinematic populations of a sample of galaxies, it is more meaningful to consider a Bayesian approach such as that from \cite{vandeSande2021statistics}. 
For the purposes of this work, however, a simple cut allows us to demonstrate the impact of resolution on the construction of mock observations. 
Hence, when the following conditions based on the observed spin parameter and ellipticity of the object, $\varepsilon$, as measured at R$_e$ are met, the object is classified as a SR:
\begin{align}
    \lambda_R &< 0.08 + \varepsilon / 4, \\
    \varepsilon &< 0.4.
\end{align}

Commonly also used in the literature to quantify the level of rotation versus dispersion support, we measure the stellar $V / \sigma$ within R$_e$ \citep{Cappellari2007SAURON, Emsellem2007Saurongalaxies}, as calculated using,
\begin{equation}    
    V / \sigma = \frac{\sum_{i=1}^{n_b} \sqrt{F_i V_i^2}}{\sum_{i=1}^{n_b} \sqrt{F_i \sigma_i^2}},
    \label{eq:vsigma}
\end{equation}
where the same definitions are used in Equation \ref{eq:lr}.

Finally, we also measure the projected stellar angular momentum, $j_*$, within R$_e$ \citep{Cortese2016SAMIangularmom} using,
\begin{equation}
    j_* = \frac{\sum_{i=1}^{n_b} F_i R_i |V_i|}{\sum_{i=1}^{n_b} F_i},
    \label{eq:j*}
\end{equation}
where, again, we use the same variable definitions as in Equations \ref{eq:lr}
and \ref{eq:vsigma}.

Having introduced our models, outlined how our mock observations are made, and defined the kinematic parameters that are measured from these observations, we can now explain how the convergence test has been performed.

\section{The convergence test}
\label{sec:convergence}

\begin{figure}
    \centering
    \includegraphics[width=\linewidth]{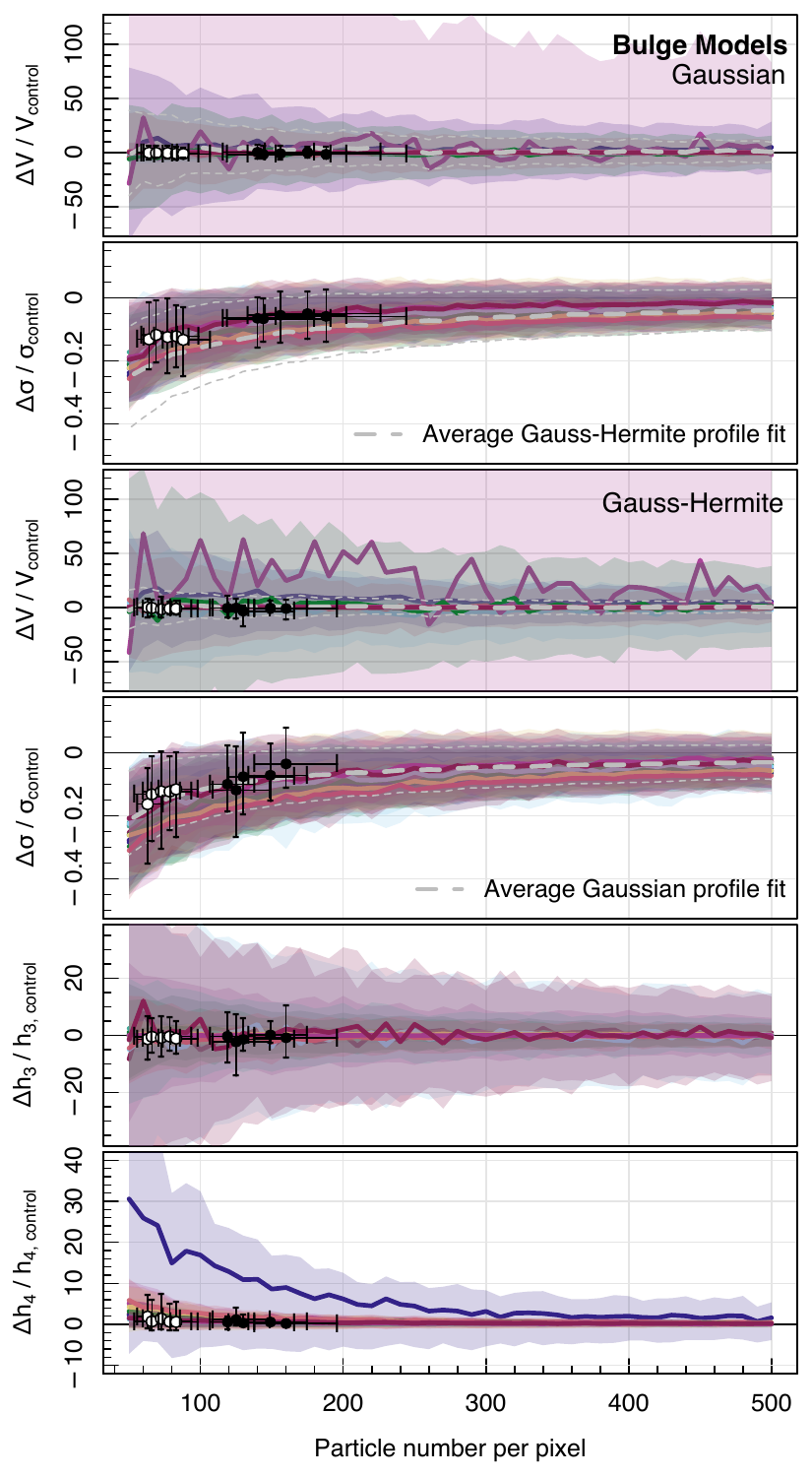}
    \caption{Results of the convergence test performed for the idealised bulge $N$-body model. From top to bottom, we examine the relative differences between the \texttt{control} model and the measured kinematic parameter from a two-moment Gaussian fit, $V$ and $\sigma$, and the four-moment Gauss-Hermite fit $V$, $\sigma$, $h_3$ and $h_4$ e.g. $\Delta V = (V_{\text{sampled}}$ - $V_{\text{true}}$) / $V_{\text{true}}$ for increasing numbers of particles per pixel. Each line traces the bootstrap distribution for a single pixel of 16, with the solid line showing the median and the shaded region showing the interquartile range of of 500 draws for that number of particles at that pixel. Points show the median and interquartile range of relative difference values for the same set of pixels within models of increasing resolution \texttt{p0}, \texttt{p16}, \texttt{p50}, \texttt{p84}, \texttt{p100} and the \texttt{control} model. The white points give the summary for outer pixels from the mock image and black points give summary of inner pixels. The same selection of pixels are used throughout the different resolution models. The grey dashed line in the top four figures shows a comparison between the average uncertainties across all 16 pixels in the fit using the \textit{other} LOSVD parameterisation (i.e. allowing comparison between a Gaussian and Gauss-Hermite recovery of $V$ and $\sigma$).} 
    \label{fig:bootstrap_bulge}
\end{figure}

\begin{figure}
    \centering
    \includegraphics[width=\linewidth]{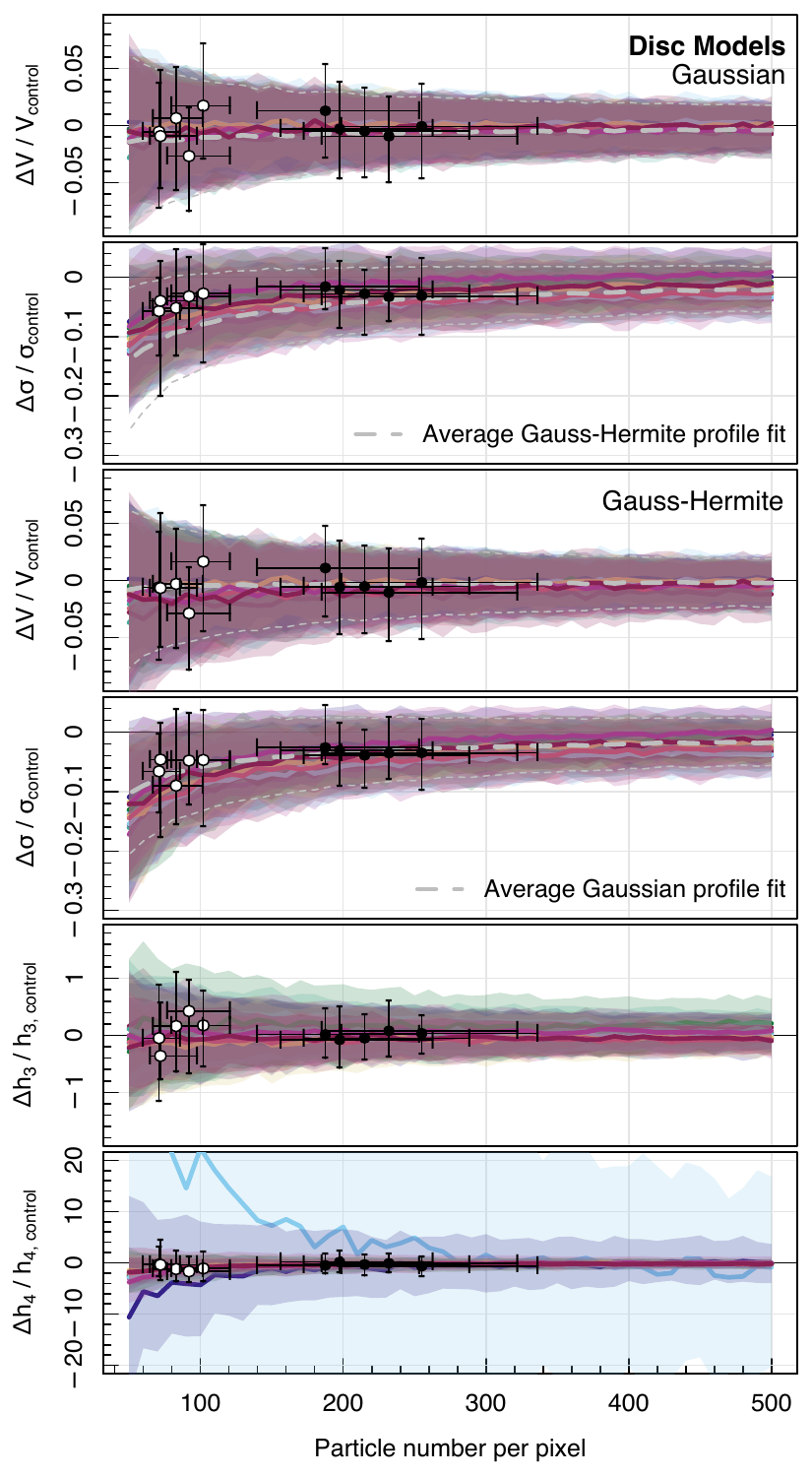}
    \caption{As in Figure \ref{fig:bootstrap_bulge}, but this time for the idealised disc $N$-body model. We show the relative residual kinematic measurements from a two-moment Gaussian fit, $V$, $\sigma$, and the four-moment Gauss-Hermite fit $V$, $\sigma$, $h_3$, $h_4$ in each panel from top to bottom. Individual colours denote a different pixel in the mock observation, were we have run 16 pixels in total. Each solid line shows the median from 500 draws made for that number of sampled particles per pixel. The shaded region shows the interquartile range of residuals in that pixel. Points and associated error bars show the median and interquartile range relative differences for mock SAMI observations of the resolution suite models in comparison to the \texttt{control} model. White points refer to the same selection of outer pixels in the image, while black points refer to the consistent selection of inner pixels. The grey dashed line in the top four figures shows a comparison between the average uncertainties across all 16 pixels in the fit using the \textit{other} LOSVD parameterisation (i.e. allowing comparison between a Gaussian and Gauss-Hermite recovery of $V$ and $\sigma$). }
    \label{fig:bootstrap_disk}
\end{figure}

It is necessary to begin with an assumption. 
We assume that the \texttt{control} models of a bulge and disc are sufficiently well-resolved to accurately recover the underlying velocity distribution. 
From Figures \ref{fig:bulge_cumulative_mass} and \ref{fig:disk_cumulative_mass}, this assumption seems a reasonable one, as even when using a fraction of the number of particles, we recover the mass and velocity distributions within 1 R$_{e}$. 
To explore how measured properties in a given spaxel change with the number of particles in that bin, we can sample this underlying ``true'' distribution to determine the cut off below which shot noise dominates our measurement. 

We do this by building a mock observation of the \texttt{control} simulation, one for the bulge and one for the disc, focusing on just 16 pixels where the number of particles per pixel exceeds 800.
In each of these pixels, we then randomly sample a fixed number of particles from that pixel in the \texttt{control} image, replacing each particle as we go. 
This is done from 50 to 500 particles, increasing by a sequence of 10 sampled per pixel. 
This bootstrap sampling process is repeated 500 times for each pixel and the LOSVD is fit and the kinematic parameters, $V$, $\sigma$, $h_3$ and $h_4$, using the Gauss-Hermite function described in Equation \ref{eq:LOSVD}, are recorded in each case. 
We then repeat this process again, independently fitting the Gaussian function described in Equation \ref{eq:Gaussian} and record the kinematic parameters $V$ and $\sigma$ from this two component fit.
We then consider the relative difference between the sampled kinematics measured and the assumed ``true'' kinematics of the \texttt{control} pixel (i.e. ($V_{\text{sampled}}$ - $V_{\text{true}}$) / $V_{\text{true}}$) and compute the mean and interquartile spread of these relative differences across 500 draws.

The aim of this convergence test is to explore how the measured properties in a given spaxel change with the number of particles in that bin. 
The resulting differences between the kinematics are shown in Figures \ref{fig:bootstrap_bulge} and \ref{fig:bootstrap_disk} for the bulge and disc \texttt{control} tests respectively. 
Each coloured line tracks the change in kinematics measured for a single of the 16 pixels.
The shaded region around each solid median shows the interquartile distribution of the residual kinematics in that pixel. 
Kinematics, from top to bottom, are shown for $V$ and $\sigma$ as fit with the Gaussian, followed by $V$, $\sigma$, $h_3$ and $h_4$ as fit using the Gauss-Hermite function with increasing particles per pixel. 

It is worth noting that, in the case of the bulge model shown in Figure \ref{fig:bootstrap_bulge}, the raw measured $V$ and higher-order kinematics, $h_3$ and $h_4$, are very small values ($V$ = [-8, 16] km/s, $h_3$ = [-0.06, 0.04], $h_4$ = [0.01, 0.2]).
Due to these small numbers, small variations show large relative differences that, on average, reduce with increasing numbers of particles. 
There are cases in the bulge model, such as the mauve pixel line shown for the $\Delta V$ and the purple pixel line shown for $\Delta h_4$, where we see variation exists even at high particle sampling. 
In these cases, the value in the control pixel is very small (in the mauve pixel, $V = 0.028$ km/s; in the purple pixel, $h_4 = 0.004$), small variations cause large residuals.   
The vast majority of pixels do show a significant narrowing of the residuals as the number of particles per pixel increases.

By comparison, the disc model results shown in Figure \ref{fig:bootstrap_disk} show much smaller relative residuals, as expected for the $V$ values where the dynamic range of measurements is more substantial ($V$ = [107, 120] km/s). 
We are no longer in a small number regime for the majority of cases, other than the higher-order moment $h_4$, where the blue pixel has a value $h_4 = 0.001$.
Importantly, again we see the scatter about a zero residual significantly decrease with increasing particles number. 
The $\Delta$ values never reduce to zero scatter even at the highest numbers of particles seen in Figures \ref{fig:bootstrap_bulge} \& \ref{fig:bootstrap_disk}. 
This is an expected result of shot noise, scaling as $1/\sqrt{N}$, but never fully reaching zero; this cannot be entirely overcome as our galaxies are discretised particle models binned into discrete pixels along both the spatial and velocity axes. We cannot overcome the impact of noise completely but here attempt to parameterise when it becomes dominant.

When comparing the residuals of the Gaussian and Gauss-Hermite fit, we see a reduction in the average relative offsets between $\sigma$ measures in particular, though the difference is small. 
This is prudent to note given consideration of methodology behind the construction of mock observations, whether kinematics quoted are 4-moment or 2-moment fits.

We further explore these trends using the range of particle resolution models: \texttt{p0}, \texttt{p16}, \texttt{p50}, \texttt{p84}, \texttt{p100}. 
For this experiment, we take a mock observation of the full galaxy using the \simspin{} SAMI instrument with the objects projected at the same fixed distance and projection angle. 
This is done for all of the simulations in the suite, including the \texttt{control} model.
We then examine how the observed kinematics change in a selection of pixels within the inner and outer regions (shown by black and white points respectively in Figures \ref{fig:bootstrap_bulge} and \ref{fig:bootstrap_disk}).
The regions are divided in this way so that we can see the variation in the residual kinematics within similar particle number density locations. 
``Inner'' and ``outer'' regions are defined by dividing pixels into two groups based on the number of particles per pixel contained in the lowest resolution model \texttt{p0}. 
The inner region is defined by pixels containing more than 100 particles, while outer region is those pixels with fewer than 100, but more than 50. 
The same pixels are then used in each subsequent image of models increasing in resolution and residuals are measured relative to those same pixels in the \texttt{control} observation. 
The error bars show the interquartile range of particle count per pixel and the residual kinematics. 
We see that the results from the different models agree closely with the bootstrap sampled distributions, generally sitting closer to the true kinematics, but with increasingly tall error bars as the number of particles per pixel reduces and shot noise begins to dominate the measured kinematics. 

\begin{figure*}
    \centering
    \includegraphics[width=\linewidth]{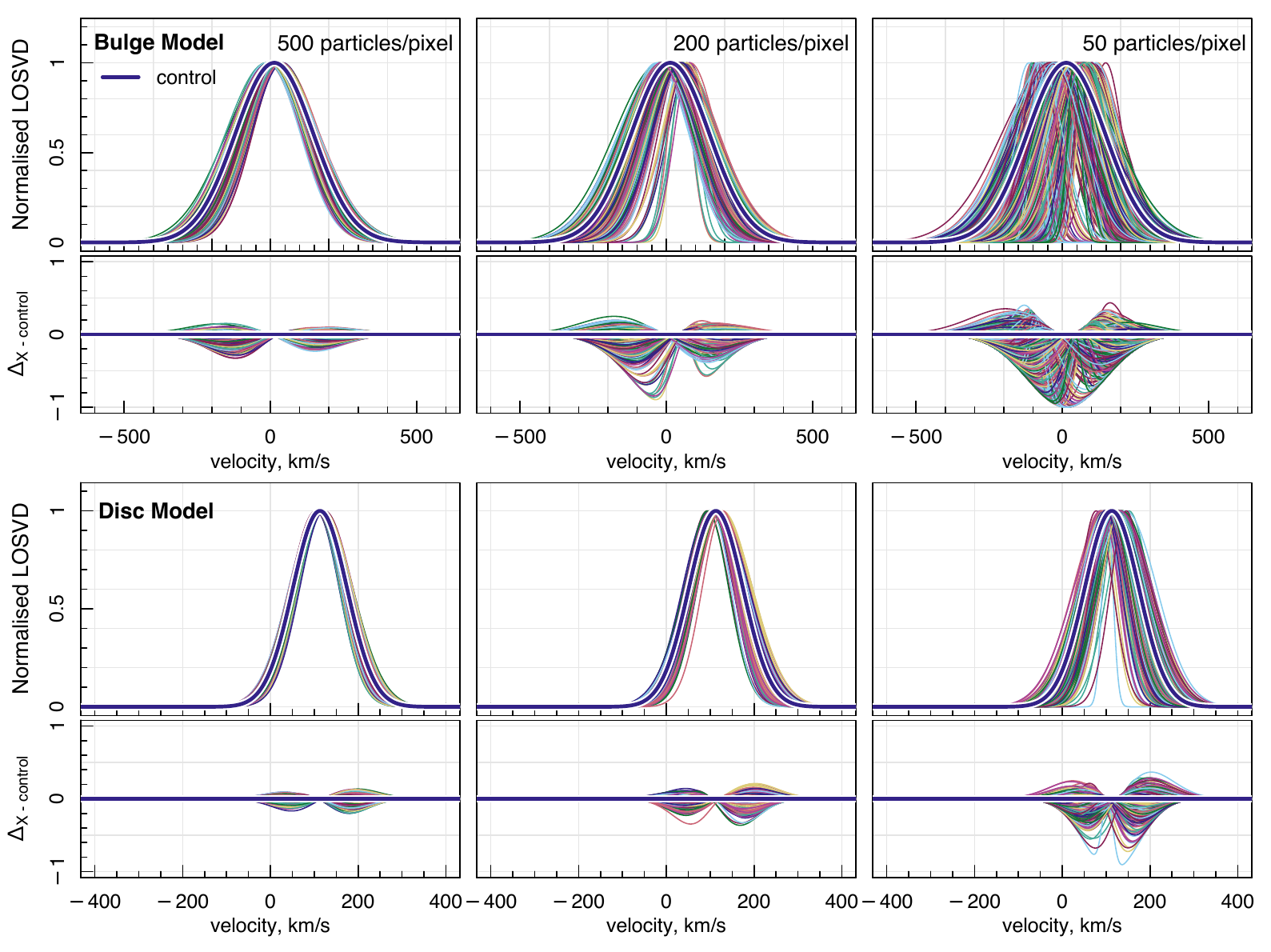}
    \caption{Considering the LOSVD within a single pixel of the bootstrap experiment with their corresponding residuals $\Delta$ LOSVD$_{\text{control}}$ - LOSVD$_{\text{bootstrap}}$ . (\textit{Panels above}) The control LOSVD in comparison to 500 bootstrap samples of a single pixel from the bulge model analysis. (\textit{Panels below})  The control LOSVD in comparison to 500 bootstrap samples of a single pixel from the  disc model analysis. The solid purple lines show the LOSVD measured in the control pixel, while the other coloured lines show the 500 bootstrap samples from that same pixel with 500 particles (\textit{left}), 200 particles (\textit{centre}) and 50 particles (\textit{right}). Clearly the distribution is more uncertain with lower numbers of draws/particles and the $\sigma$ of the distribution is systematically under-represented in sampled draws as seen by the predominantly negative residuals below each LOSVD. Further clearly demonstrated is that the number of required particles varies with the underlying kinematic morphology.}
    \label{fig:bootstrap_losvd_summary}
\end{figure*}

In Figure \ref{fig:bootstrap_losvd_summary}, we present the bootstrap sampled LOSVD for a single pixel from each model for the 500, 200 and 50 particle per pixel cases. 
For each panel, we plot the \texttt{control} model LOSVD as a purple line with white edging for clarity and consider the residual of the \texttt{control} minus the sampled pixel LOSVD for each scenario.
Behind each line, we plot each LOSVD of 500 bootstrap sample of (500/200/50) particles from that pixel.
The increase in noise in lower particles per pixel samples are evident. Similarly, the underestimation of the dispersion, $\sigma$ is clear from the overly negative residuals.
Finally, in comparing the top and bottom panels for the bulge and disc simulation respectively, we see the importance of testing both extremes of the kinematic morphology distribution.
For systems with higher dispersion support (i.e. the bulge model), greater numbers of particles are needed to achieve a similar level of noise in the resulting LOSVD.

Consistency between the bootstrap test and models of different resolution fuel the importance of setting limits on the number of particles per pixel required to recover observable kinematics. 
The resolution models have similar particle numbers to what you would expect to draw from a galaxy formation simulation, such as \eagle{}. 
This demonstrates where the typical system from such a simulation will sit in this parameter space, and we can clearly see that, especially in the case of the galaxies with dominant dispersion components such as a bulge, the recovery of the LOS dispersion is always likely to be under-estimated. 

We can turn these results around and use them to answer: how many particles per pixel should you aim for in a given mock observation? 
For the bulge and disc models separately, summarising over all pixels, we compute the 2-sigma distribution of 500 kinematic residuals at each of the particle per pixel sampling bins from 50 to 500. 
This tells us the accuracy within which we can expect 95\% of our data to lie.
We compute a fit to this distribution and determine the point at which increasing the number of particles per pixel will no longer be of benefit relative to some intrinsic uncertainty associated with the measurement itself.
These results are summarised in Table \ref{tab:uncertainty} for both 2- and 4-moment LOSVD fits.
The empirical equations used to determine these values are outlined in Appendix \ref{app:uncertainty} such that the reader can determine the number of particles for any required certainty bounds.  

\begin{table}
\caption{Considering the number of particles required to reach a $2\sigma$-uncertainty of a given size on each of the measured orders of the LOSVD for a dispersion dominated and rotation dominated galaxy. Values given with a $*$ denote those values which have been extrapolated given the fitted trends from our bootstrap test.}
\begin{tabular}{@{}|l|c|cc|@{}}
\toprule
\multirow{2}{*}{\textbf{\begin{tabular}[c]{@{}l@{}}Kinematic \\ Parameter\end{tabular}}} & \multirow{2}{*}{\textbf{\begin{tabular}[c]{@{}c@{}}$2 \sigma$-\\ uncertainty\end{tabular}}} & \multicolumn{2}{c|}{\textbf{Number of Particles}} \\ \cmidrule(l){3-4} 
 &  & \multicolumn{1}{c|}{\begin{tabular}[c]{@{}c@{}}Dispersion \\ Dominated\end{tabular}} & \begin{tabular}[c]{@{}c@{}}Rotation \\ Dominated\end{tabular} \\ \midrule
$V_{\text{2-mom}}$, km/s & 20 & \multicolumn{1}{c|}{217} & 67 \\
 & 15 & \multicolumn{1}{c|}{373} & 114 \\
 & 10 & \multicolumn{1}{c|}{792*} & 248 \\ \midrule
$\sigma_{\text{2-mom}}$, km/s & 30 & \multicolumn{1}{c|}{136} & 23 \\
 & 20 & \multicolumn{1}{c|}{337} & 49 \\
 & 10 & \multicolumn{1}{c|}{6771*} & 231 \\ \midrule
$V_{\text{4-mom}}$, km/s & 20 & \multicolumn{1}{c|}{270} & 81 \\
 & 15 & \multicolumn{1}{c|}{436} & 134 \\
 & 10 & \multicolumn{1}{c|}{812*} & 279 \\ \midrule
$\sigma_{\text{4-mom}}$, km/s & 30 & \multicolumn{1}{c|}{206} & 23 \\
 & 20 & \multicolumn{1}{c|}{535*} & 54 \\
 & 10 & \multicolumn{1}{c|}{1858*} & 241 \\ \midrule
$h_{3}$ & 0.15 & \multicolumn{1}{c|}{149} & 122 \\
 & 0.1 & \multicolumn{1}{c|}{264} & 242 \\
 & 0.05 & \multicolumn{1}{c|}{829*} & 809* \\ \midrule
$h_{4}$ & 0.15 & \multicolumn{1}{c|}{271} & 140 \\
 & 0.1 & \multicolumn{1}{c|}{446} & 242 \\
 & 0.05 & \multicolumn{1}{c|}{861*} & 613* \\ \bottomrule
\end{tabular}
\label{tab:uncertainty}
\end{table}

\begin{figure*}
    \centering
    \includegraphics[width=\linewidth]{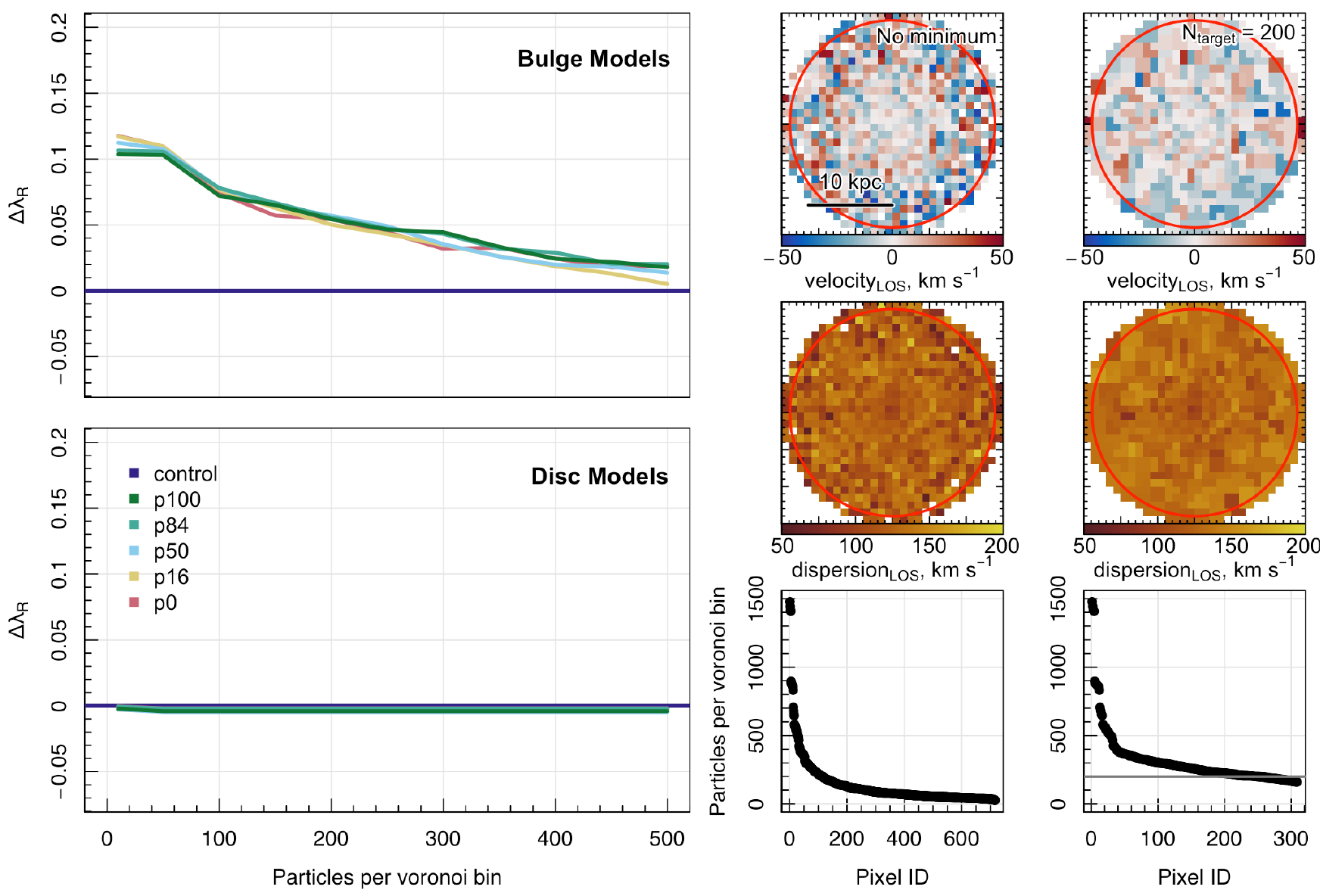}
    \caption{Considering the impact of measuring the spin parameter, $\lambda_R$, within mock observations of increasing Voronoi-binned pixels to meet some minimum number of particles-per-pixel. (Left) We consider a summary of this experiment. We measure $\lambda_R$ from SAMI observations of the \texttt{control} $N$-body models of bulge (above) and disc galaxies (below), as shown by the purple line, and compared to the \texttt{p0}, \texttt{p16}, \texttt{p50}, \texttt{p84} and \texttt{p100} models in the other colours from pink to green. All $\lambda_R$ values are measured within the same effective radius, $R_{\text{eff}}$, as shown by the red ellipse region in the mock observations on the right. The difference between $\lambda_R$ measured in the lower resolution models and the \texttt{control} models are shown as a raw difference ($\Delta \lambda_R = \lambda_{R, \texttt{pX}} - \lambda_{R, \texttt{control}}$). We can see that, with insufficient numbers of particle per pixel, the $\lambda_R$ measured for dispersion dominated systems in always overestimated. (Right) We demonstrate the result of using the voronoi binning algorithm described in \S \ref{sec:voronoi_bin} as run for the \texttt{p0} bulge model. The inner column of panels demonstrate the LOS velocity and dispersion maps for the dispersion-dominant model when no binning is implemented. In the lowest panel of this column, we show the number of particles per pixel. The final column shows the voronoi binned images of the same dispersion-dominant model built to a target number $N_{\text{target}} = 300$. The effect of this minimum target is demonstrated by the change in pixel shape and the increase in the minimum number of particles per bin shown in the lowest panel.}
    \label{fig:nbody_voronoi}
\end{figure*}

\section{Implications}
\label{sec:implications}

\begin{figure*}
    \centering
    \includegraphics[width=\linewidth]{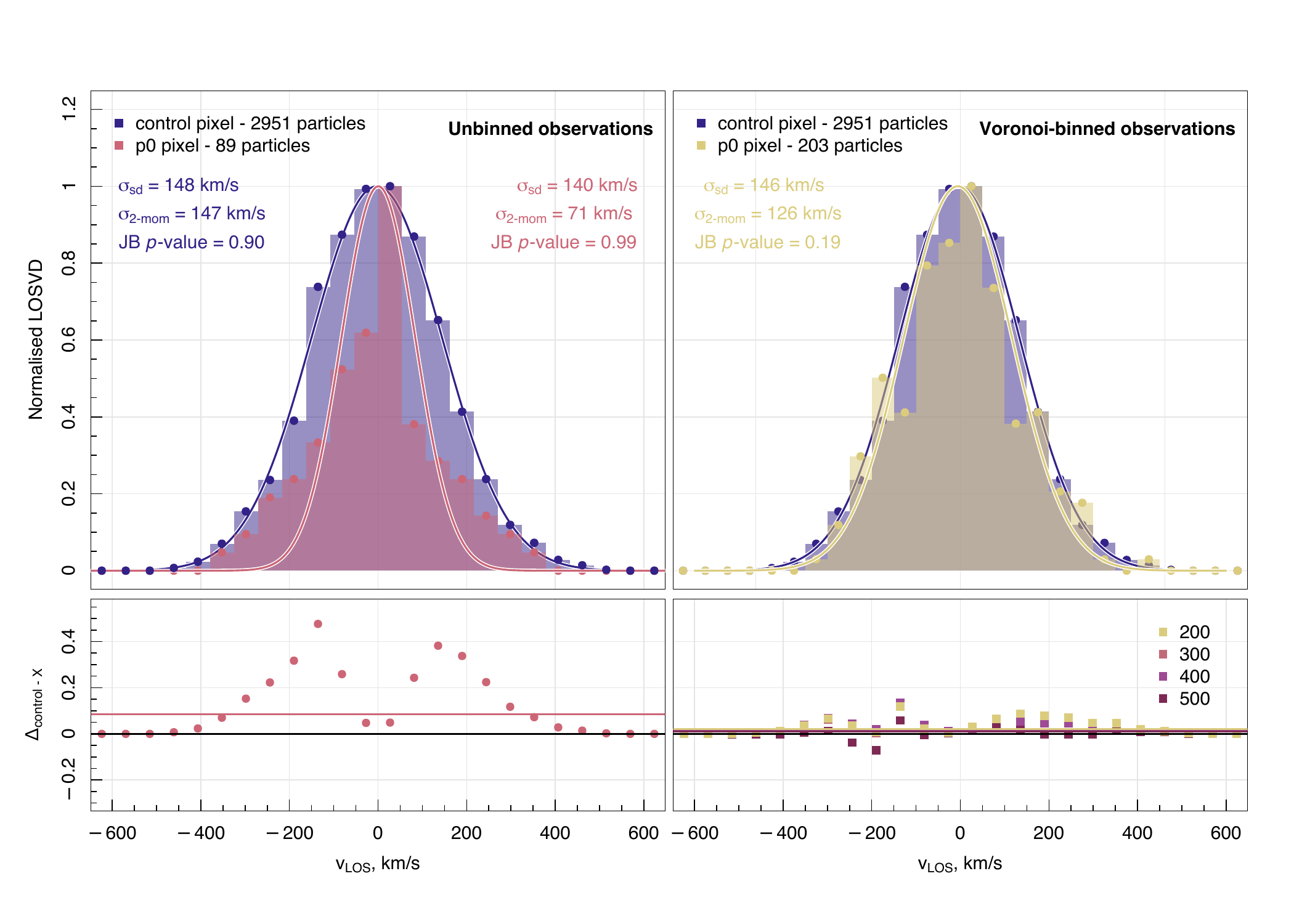}
    \caption{Showing the normalised LOSVD for unbinned observations (upper left) and Voronoi-binned observations (upper right) for a single consistent pixel in the control N-body bulge model versus the \texttt{p0} model. We quote the Jarque-Bera (JB) test \textit{p}-value for each distribution and find all distributions are compatible with normality. The residual between the LOSVD of the \texttt{p0} model and the high-resolution control (lower panels). On the left, we see the \texttt{p0} LOSVD is not well approximated by a Gaussian, with the fitted 2-moment Gaussian distribution appearing narrower ($\sigma_{\text{2-mom}} = 71$ km/s) than that seen in the control model ($\sigma_{\text{2-mom}} = 147$ km/s). The residuals in the panel below demonstrate the magnitude of this difference, where the horizontal pink line shows the standard deviation in the spread of residuals. On the right, we show the same pixel but in this case for Voronoi-binned observations. The LOSVD's are shown for the original control pixel and the same pixel in the \texttt{p0} model when Voronoi binned to > 200 particles ($\sigma_{\text{2-mom}} = 126$ km/s). In the residuals below we show the differences between the LOSVD's in increasingly Voronoi binned pixels from 200-500. We clearly see that, even binning to just 200 particles per pixel, we significantly reduce the difference between our fitted LOSVD in comparison to the high-resolution control model. We also show the standard deviation ($\sigma_{\text{sd}}$) measured for all particles in the pixel in each case.}
    \label{fig:losvd_voronoi}
\end{figure*}

In the scenario that a model does not have enough particles-per-pixel within the desired observational set-up requirements, is there anything we can do? 
We have demonstrated where \eagle{} simulations would sit within the kinematic uncertainty parameter space and although the resolution parameters are generally well-matched in gravitational softening and physical resolution to the observational set-up of SAMI, they will appear biased due to this particle-per-pixel density. 
In particular, spheroidal objects with dominant dispersion-supported structures will be affected more strongly than their rotation-supported counterparts, creating a systematic bias in the way we visualise the spin-ellipticity plane. 
This has an effect on the fraction of slow rotators observed in a given simulated Universe, and as this is used as a method of comparing different observational surveys and theoretical models, it's greatly important that this systematic bias is accounted for. 

Using an approach similar to that of observations in tackling signal-to-noise effects, here we suggest Voronoi binning methods to increase the number of particles per pixel within which the kinematics are measured. 
This has been implemented in an upgrade to the \simspin{} package v2.7.1, which we describe in Section \ref{sec:voronoi_bin}.
We follow this in \S\ref{sec:previous_work} by a discussion of how these results may impact previous work in the literature that consider the slow-rotator fraction of modern galaxy formation simulations using mock observations. 

\subsection{On future measurements}
\label{sec:voronoi_bin}

Voronoi tessellation is an adaptive binning technique used to maintain a uniformity in a specified property across the image. 
In the observational case, this conserved property was signal-to-noise (S/N).
Implemented for integral field spectroscopy data, \cite{Cappellari2003Vorbin} applied this method such that a greater number of pixels within the observation could be used for science by binning pixels to meet some minimum S/N requirement. 
Pixels that do not meet a minimum threshold before binning are grouped into larger bins until that threshold is met. 
This is done adaptively across the image such that bright regions at the centre that already meet the minimum S/N threshold are not binned (hence maintaining the spatial resolution of the instrument) while outer pixels could be grown until they are suitable for kinematic fitting.
For the theory data, we follow a similar approach.
In this case, we bin pixels to meet some minimum number of particles-per-bin, rather than some S/N value. 

Further algorithmic requirements are considered in the Voronoi tessellation procedure.
These are fully detailed in the work of \cite{Cappellari2003Vorbin}, but we highlight the key points here.
For example, when growing bins to meet some minimum number of particles-per-pixel, we require that bins must be (1) spatially connected, (2) uniform and (3) round. 
These three requirements prevent bins from growing particularly long, thin or large, such that the spatial resolution is maintained as well as possible. 
The rules also prevent pixels from becoming isolated and unable to be binned as a result. 

In v2.7.1 of \simspin{,} we have implemented an optional Voronoi binning method for grouping individual pixels into bins that meet some minimum number of particles-per-pixel. 
This is done in the new \texttt{voronoi} function, which is called within the method for building a mock IFS data cube. 
This method can be activated by specifying the boolean parameter \texttt{voronoi\_bin = TRUE} and specifying the necessary particle number limit \texttt{vorbin\_limit} = $N_{\text{target}}$ within the \texttt{build\_datacube} function. 
Algorithmically, we follow the method below:
\begin{enumerate}
    \item Pixels that exceed the requested particle limit are assigned to bins of single pixel area. 
    \item Moving on to pixels that do not meet the minimum particle limit, we order the pixels by the number of particles contained from largest to smallest. The pixel with the greatest number of particles becomes the target pixel for binning.
    \item The distance between the target pixel and any remaining unbinned spaxels is computed and any unconnected pixels are discarded. 
    \item Between the remaining candidate pixels, we next compute how round each bin will be with the addition of the target pixel. The roundness is calculated by considering 
    \begin{equation}
    \label{eq:roundness}
        R = \frac{ \sqrt{ \sum_{i = 0,1}{\left( x_{i, \text{target pixel}} - x_{i, \text{unbinned pixel}}   \right)^2 }}  }{ \sqrt{(A_{\text{target pixel}} + A_{\text{unbinned pixel}})/\pi}} - 1,
    \end{equation}
    where $A$ is the area of the pixel and $x_i$ denotes the position in 2D space of the centroid of each pixel in the index 1 and 2 dimensions respectfully, i.e again using the distance between the target and unbinned pixels. A value of $R = 0$ will result in a perfectly round circular bin. We require that any added bins must have an $R \leq 0.3$ and discard any candidate pixels that do not meet this requirement.
    \item Finally, we consider the sum of the number of particles in the target pixel along with the remaining unbinned candidate pixels. We then select the most uniform option, where the uniformity criterion is considered,
    \begin{equation}
    \label{eq:uniformity}
        U = \left| 1 - \frac{(N_{\text{target pixel}} + N_{\text{unbinned pixel}})}{N_{\text{target}}}\right|,
    \end{equation}
     where $N$ denotes the number of particles within that pixel and $N_{\text{target}}$ denotes the target number of particles per bin. This uniformity value, $U$, is minimised. 
     \item If an unbinned pixel is remaining following each of these steps, this pixel is joined to the target to form a bin. If the minimum value of particles-per-pixel is met with this join, the bin will join a list of successfully binned pixels. If it does not yet reach that minimum particles-per-pixel value, this newly joined pixel is returned to the list of unbinned pixels for joining in the next round of the algorithm.
     \item This process is repeated until all possible pixels have been joined to a bin that meets the minimum number of particles requirement. In the case that a pixel cannot be binned, it is joined to the closest bin centre.
\end{enumerate}

This results in images such as those shown to the right side of Figure \ref{fig:nbody_voronoi}.  
Single pixels in an unbinned kinematic map of the bulge $N$-body model \texttt{p0} are grouped together until a single Voronoi bin contains within 80\% of the requested particle limit ($N_{\text{target}} = 300$ in right-hand column of Fig \ref{fig:nbody_voronoi}). 
Some wiggle room is allowed within the uniformity criterion in Equation \ref{eq:uniformity} and once bins contain at least 80\% of the requested limit, we stop the bin accretion algorithm. 
This is consistent with the methodology of \cite{Cappellari2003Vorbin}, as the alternative is that the bins will generally overshoot the limit.
In the lower panels on the right hand side of Figure \ref{fig:nbody_voronoi}, we show how the number of particles per bin is lifted per pixel following this algorithm completion. 

On the left hand side of Figure \ref{fig:nbody_voronoi}, we show how the measured $\lambda_R$ changes as a function of particles-per-pixel when Voronoi binning the mock observations of our resolution models \texttt{p0}, \texttt{p16}, \texttt{p50}, \texttt{p84}, and \texttt{p100} in comparison to the value measured from the \texttt{control} models for the bulge and disc flavours respectively in the top and bottom panels. 
We measure $\lambda_R$ in each set of mock observations within a consistent measurement radius, specifically the effective radius.
In the panels on the right of Figure \ref{fig:nbody_voronoi}, we show this measurement radius for the bulge $N$-body model as a red ellipse. 
A similar half-light ellipse is used for the disc $N$-body model, though this region is far more elliptical than the similar region for the bulge. 

When measuring $\lambda_R$ within the effective radius, R$_{e}$ we use an un-binned flux map and consider all un-binned pixels whose centre falls within R$_{e}$. 
These same pixels are then taken from the Voronoi binned velocity and dispersion maps, where any pixels that fall in the same Voronoi bin have been given the same value. 
This method reflects that done in \cite{vandeSande2019SAMISimulations}.
In each case, we plot the raw difference between the $\lambda_R$ measured within the lower resolution model and that in the \texttt{control} model, $\Delta \lambda_R = \lambda_{R, \texttt{pX}} - \lambda_{R, \texttt{control}}$.  

As can be seen in Figure \ref{fig:nbody_voronoi}, the difference between the \texttt{control} measured $\lambda_R$ and that measured within the Voronoi binned data decreases steadily as we approach 500 particles per pixel. 
There is a flattening of the distribution at $\sim 200$ particles per bin.
This distribution shows that the value of $\lambda_R$ measured in lower resolution models is always an over-estimate for systems with dispersion dominant kinematics.
The relationship is inverted with respect to the $\sigma_{\text{LOS}}$ measured in the bootstrap experiment, as shown in Figure \ref{fig:bootstrap_bulge}, where the values are always underestimated in lower particle-number pixels. 
This makes sense, as the dispersion is measured in the denominator of $\lambda_R$, as shown in Equation \ref{eq:lr}.
Approximately $90\%$ of this over-estimation of $\lambda_R$ is driven by the underestimation of $\sigma$, as determined by comparing the fractional changes in each parameter. Any remaining change is driven by noise in the measured velocity.

In Figure \ref{fig:losvd_voronoi}, we show the effect of Voronoi-binning on the raw LOSVD for a single, consistent pixel extracted from observations of the $N$-body bulge model. 
In this pixel, we plot the velocity distribution recorded in one pixel of the SAMI-like observation for the control bulge model in comparison to the \texttt{p0} model. 
At this lowest resolution, shown by the \texttt{p0} pink line on the upper left of the figure, we can see distribution is not particularly Gaussian, and the subsequent fit vastly under-estimates the dispersion of the profile ($\sigma_{\text{2-mom}} = 71$ km/s, in comparison to $\sigma_{\text{2-mom}} = 147$ km/s for the control). 
There are 89 particles in this \texttt{p0} pixel, in comparison to the 2951 particles within the \texttt{control}. 
We note that both distributions in this case are indistinguishable from normal, as given by the Jarque-Bera (JB) test for normality\footnote{We use the \textsc{R}-package `tseries' to perform this statistical test: \url{https://cran.r-project.org/package=tseries}} with both $p$-values (0.90 for \texttt{control} and 0.99 for \texttt{p0}) well above the null hypothesis limit of 0.05.
By comparison, on the right we consider the same model (the \texttt{p0} bulge $N$-body model) but this time Voronoi-binned with increasing numbers of particle-per-pixel. 
The LOSVD for the consistent pixel binned to meet 200 particles-per-pixel is shown in the upper right panel. 
Here, we see that the LOSVD is much closer to that of the control model ($\sigma_{\text{2-mom}} = 126$ km/s), with the residual difference between this and any of the increasingly binned models now sitting at an average of $\sim$ 3\% as shown by the horizontal lines in the lower right panel. 
The voronoi-binned distribution is also well described by a normal, as seen by the JB $p$-value of 0.19 (as $p$ > 0.05).
We note that, when measuring the standard deviation of this pixel's particle velocities directly ($\sigma_{\text{sd}}$), the \texttt{p0} model difference from the \texttt{control} is effectively negligible. 
However, in this example, when producing ``mock'' observations as often done in the literature, the fitting of a Gaussian distribution does have a noticeable impact on the recovered $\sigma_{\text{2-mom}}$.

Of course, the example $N$-body models shown in Figure \ref{fig:nbody_voronoi} and \ref{fig:losvd_voronoi} are taken as the extremes of the distribution of disc to bulge dominated systems. 
Within a galaxy formation simulation, we are far more likely to have a galaxy model that is a mix of the two flavours - some proportion being a rotational-dominated disc-like systems and others being dispersion-dominated elliptical objects, but with significant overlap. 
With this in mind, we consider how important such particle-per-pixel limits might be for measuring the distribution of kinematic parameters within a full cosmological volume.

Returning to the \eagle{} simulation and the \texttt{RefL0050N0752} volume, we re-run the N mock observations made of galaxies above $10^{10}$ \msol{}, implementing the Voronoi binning method for bins of $N_{\text{target}} = 200, 300$, $400$ and $500$. 
In each case, we measure the kinematics of the system within the same half-light radius, at the same $70^{\circ}$ inclination, and at the same projected distance where the semi-major axis of this measurement ellipse will fill the central 28 pixels of the \sami{'s} circular field-of-view. 
We follow the method outlined in \cite{vandeSande2019SAMISimulations} for measuring kinematics from observations that have been Voronoi binned. 

\begin{figure*}
    \centering
    \includegraphics[width=\linewidth]{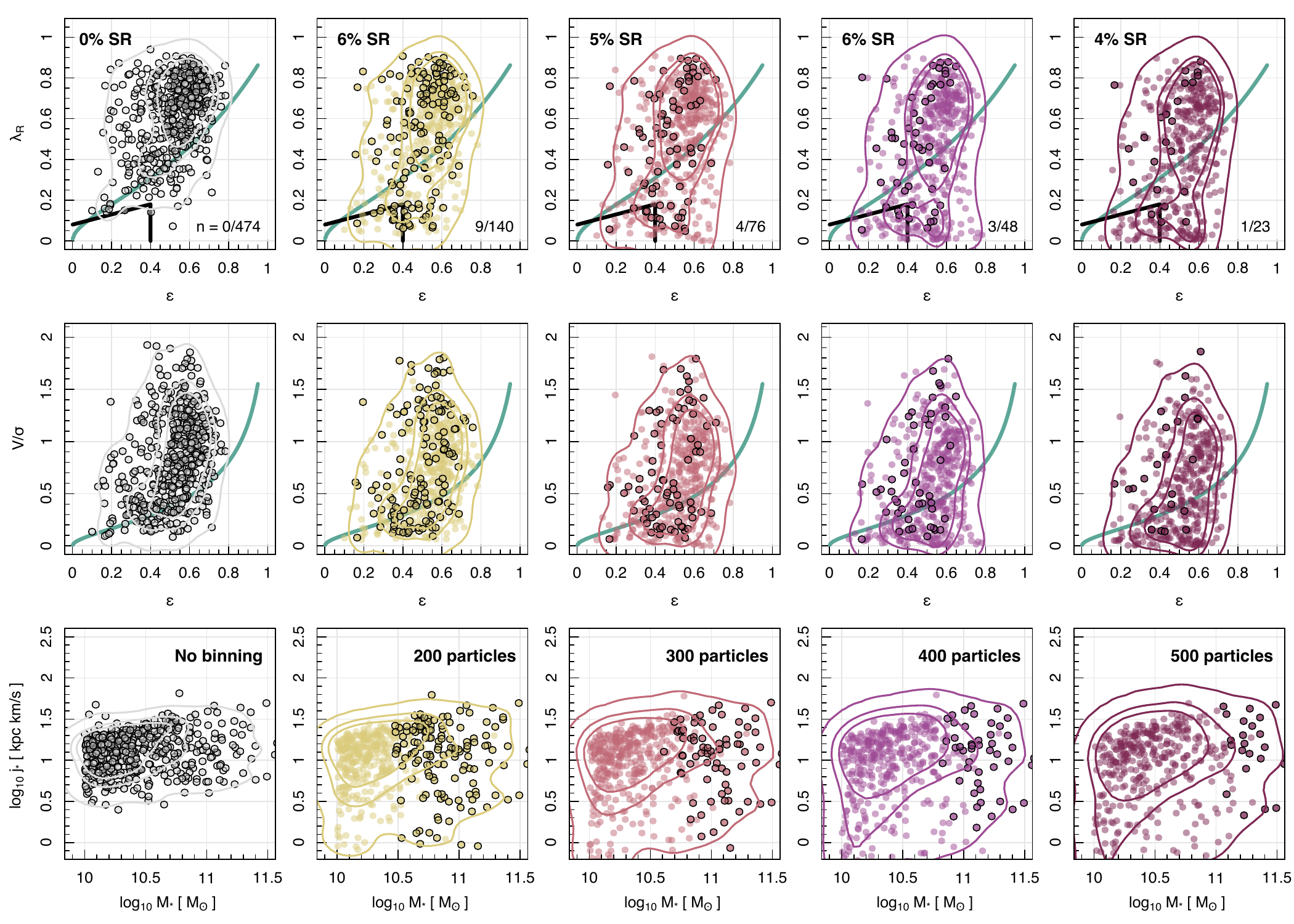}
    \caption{For galaxies extracted from the \eagle{} simulation (top row) the spin-ellipticity ($\lambda_R$-$\varepsilon$) plane, where the black line shows the slow/fast rotator boundary as suggested by \protect\cite{Cappellari2016AnnualReview} and the green line shows the theoretical position of an edge-on oblate system with given $\varepsilon$ and anisotropy described by $\beta$ = $0.7$ $\times$ $\varepsilon$; (middle row) the V$/\sigma$-$\varepsilon$ distribution with the same definition magenta line as on the left; and (bottom row) the projected stellar angular momentum $j_*$ as a function of stellar mass. Points are individual observations of galaxies extracted from \eagle{}, with those with black borders demonstrating systems with at least 50 spaxels within the measurement radius. The overlaid contours show the distribution of 95\%, 68\% and 50\% of the data. The left column in grey points show measurements made without any restrictions on the number of particles-per-pixel from which the kinematics have been measured, i.e. unbinned. The following columns show the same kinematics measured from mock images that have been Voronoi binned to meet some target number of particles-per-pixel, as listed in the bottom right hand corner of each row of images. As can be seen from the slow rotator fraction listed in the upper left of the top row, the effect of increasing the particle numbers per bin from no binning to 200 causes an increase in the fraction of slow rotators observed. We further list the (number of SRs/total number of galaxies with > 50 pixels) in the bottom right of the top row of plots. }
    \label{fig:eagle_kinematics}
\end{figure*}

Results of the \eagle{} measurements are shown in Figure \ref{fig:eagle_kinematics}. 
Each row of plots demonstrates a different binning limit. 
The left column shows the kinematics measured within 1 R$_e$ for systems that have not been binned at all. 
Subsequent columns show the results of Voronoi binning with $N_{\text{target}} = 200, 300, 400$ and $500$ particles per pixel respectively. 
In the top row, we show the $\lambda_R$-$\varepsilon$ plane, followed by the $V/\sigma$-$\varepsilon$ parameter space and the $j_*$-$M_*$ distribution in the final row.
We further measure the fraction of slow rotators in each case, using the definition of \cite{Cappellari2016AnnualReview} as shown by the black box in the lower left of the $\lambda_R-\varepsilon$ plane.

As we increase the number of particles per Voronoi bin, we begin to reduce the spatial resolution within that measurement radius, R$_e$, which presents a resolution systematic of its own. 
We show observations of all systems, but observations with 50 unique bins or more are shown with outlined points in Figure \ref{fig:eagle_kinematics}.

There is a stark difference between the column on the left and the rest of the Voronoi binned results. 
We remind the reader that the lack of slow rotators in the un-binned mock observations is a result of fixing the observed inclination at 70 deg. 
This choice allows us to disentangle the observational effects of inclination from the resolution question we wish to investigate.
Once Voronoi binning has been implemented, we see that the $f_{SR}$ stabilises between $4-6\%$, partially driven by the sample size reducing as we bin to larger and larger particle numbers. 
The number of dispersion dominated objects and number of galaxies observed in each panel with greater than 50 unique pixels within the measurement radius are shown as a fraction in bottom-left corner of the top row of panels.

In the lowest panel, it is clear that the requirement of at least 50 pixels within the measurement radius does implement an effective stellar mass limit in quoted kinematic results, though it is important to note that this will be heavily affected by the choice of instrumental spatial resolution chosen for the mock observation telescope. 
In the scenario that an instrument with smaller spaxel sizes is required, this effective mass limit would also go up as more particles per spatial resolution element would be required to meet the particle-per-pixel limit. 
It is more meaningful therefore to determine the required resolution of the mock observations by the particle-per-pixel limit than an effective cut in stellar mass.

\subsection{On previous findings}
\label{sec:previous_work}

Considering previous works that have measured the spin-ellipticity ($\lambda_R$-$\varepsilon$) plane and slow-rotator fraction ($f_{SR}$) for different simulations, we now discuss the implications of this result.
We take care to note in which cases simulated LOSVD's have been fitted using a Gaussian, a Gauss-Hermite function or computed using the mean and standard deviation of the particle velocities directly. 

In \cite{Lagos2018Thegalaxies}, kinematic maps for \eagle{} galaxies were made using a fixed pixel size of 1.5 pkpc, akin to the spatial resolution of \sami{.} 
In this scenario, LOSVD's have been fitted using a Gaussian function. 
From these maps, $\lambda_{r50}$ was measured for galaxies with stellar masses above $5 \times 10^{9.5}$ \msol{}, where $\lambda_{r50}$ was measured using pixels within the stellar half-mass radius. 
The resulting slow rotator fractions were found to qualitatively show the expected distributions as a function of stellar mass, with $f_{SR}$ increasing with stellar mass as seen in the observational results from the \sami{} \citep{vandeSande2017SAMIKinematics} and MaNGA surveys \citep{Greene2018MaNGAAMGalaxies}. 
Similarly, when considering the environment of the galaxy, as dictated by their ``central'' or ``satellite'' status, the qualitative trends were in line with observations.
The effects of shot noise on this result will not impact these conclusions.
However, there is a systematic under-estimation of these slow rotator fractions seen in the comparison between \eagle{} and \sami{,} especially at the massive end of the galaxy sample where $f_{SR}$ is seen to flatten. 
While a significant amount of this discrepancy is due to the implementation of active galactic nuclei (AGN) physics in these models as discussed in the works of \cite{Bahe2017Hydrangea} and \cite{Barnes2017CEAGLE}, some portion of this may also be impacted by the numbers of particles in each 1.5 pkpc pixel used for the mock kinematic maps from which $\lambda_{r50}$ has been measured. 
The magnitude by which AGN feedback is not strong enough in these massive systems may similarly be over-estimated as a result.

In particular, we know that dispersion dominated systems are more dominant at the high-mass end of the distribution.
This higher fraction of dispersion dominated systems is also where increasing numbers of particles are required to accurately recover the measured kinematics, which may result in some portion of the bias. 
The change in SR fraction seen in Figure \ref{fig:eagle_kinematics} varies by as much as an order of magnitude, though it is important to remember here that we have taken a very selective orientation angle ($70^{\circ}$ inclination) for each galaxy in this case, unlike in the work of \cite{Lagos2018Thegalaxies} where a random orientation is used. 

In the works of \cite{vandeSande2019SAMISimulations, vandeSande2021statistics}, the kinematics from several cosmological galaxy formation models were compared to IFS observations from the \sami{} survey. 
In these works, individual simulation teams from the \eagle{}, \textsc{HorizonAGN} and \textsc{Magneticum} simulations have constructed their own mock observations. 
In the former two simulations, a fixed pixel size of 1.5 pkpc is used to construct mock observations and a Gaussian function is used to fit the LOSVD in each, while the \textsc{Magneticum} simulation has produced Voronoi binned maps to avoid low particle numbers per pixel (N$_{\text{target}}$= 100) and compute $V_{\text{los}}$ and $\sigma_{\text{los}}$ from the mean and standard deviation of particle velocities in each bin, as in \cite{Schulze2018KinematicsMagneticum}.
The pixel sizes in this case are not tuned to match the spatial resolution of the SAMI survey. 
The \textsc{HorizonAGN} team measure their kinematics in pixels with more than 10 particles, while no lower limit is imposed on the \eagle{} measurements. 

Given the results shown in our convergence test, we would expect there to be a systematic underestimation of the measured $\sigma$, as is shown in \cite{vandeSande2019SAMISimulations}'s figure 5, in comparison to observationally matched samples. 
That being said, given the Voronoi binning implemented in the \textsc{Magneticum} simulation results, we might have expected this offset to be smaller than the other simulation results in which pixels were not binned. 
This does not appear to be the case, with all simulations seeming to under-estimate the observed distribution of $\sigma$.
However, this also may be driven by methodological differences between the construction of the \textsc{Magneticum} observations and observations of the \eagle{} and \textsc{HorizonAGN} simulations. 
Pixels are not gridded to match the spatial resolution of SAMI and as such the smoothing effect caused by this spatial binning is reduced. 
It is noted in \cite{vandeSande2019SAMISimulations} that this difference may also arise from the fact that \textsc{Magneticum} under produces the number of round galaxies observed, as shown in figure 4 of their work, perhaps being due to the smaller volume assessed by \textsc{Magneticum} in comparison to \eagle{} and \textsc{HorizonAGN}.
We would expect the higher dispersion systems to dominate in $\varepsilon < 0.2$ objects.
Coupled with the various implementations of star formation, stellar and black hole feedback, it is difficult to disentangle the primary drivers of the differences between these mocked and true observations when a consistent particle-per-pixel limit and mock observation methodology is not maintained.  

In the work of \cite{WaloMartin2020KinematicAccretion}, a lower particle-per-pixel limit of 10 is considered.
Assuming that the distribution of particles in observational maps follows a Poisson distribution, the signal-to-noise (S/N) per pixel should follow $1 / \sqrt{N}$, and given a required S/N > 3 \AA{}$^{-1}$ to recover stellar kinematics in \cite{vandeSande2017SAMIKinematics}, this explains the lower limit of 10 particles. 
As a result, the same limit is used in \cite{Sarmiento2023MaNGIAanalysis}. 
However, the approach of measuring the LOSVD in each of these cases is very different. 

Within \cite{WaloMartin2020KinematicAccretion}, a number of different methods of classifying the LOSVD are used: (1) using luminosity weighted mean and standard deviation of the particle velocities, and (2) fitting a Gaussian distribution to a velocity-binned LOSVD distribution weighted by particle luminosities. 
The latter methodology is similar to the approach followed by \simspin{} when run in ``\texttt{velocity}'' mode, where we have used the 2-moment Gaussian LOSVD fit in our presentation of integrated kinematics.
In the case of measuring the mean and standard deviation of the particle velocities as the first and second moments directly, this limit may be suitable.
However, in the case of fitting a Gaussian to the binned distribution, our work demonstrates that this will result in a significant bias for objects with significant dispsersion support. 
As a result of placing just a 10 particle-per-pixel limit, we expect that the $\sigma_{\text{LOS}}$ computed for these observations are likely underestimated.
We may expect that the steep change in the mass-size relation observed in their figure 6 at a transition mass of $10^{10.3}$ \msol{} may have a weaker dependence on the observed $\lambda_R$ parameter in the case that this systematic were controlled.

In contrast, the work of \cite{Sarmiento2023MaNGIAanalysis} assigns individual particles to spectra which are shifted to reflect the velocity of each particle, gridded onto the observed spectral wavelengths and summed to give an observed spectrum per pixel. 
This methodology is akin to the methodology of \simspin{} run in ``\texttt{spectral}'' mode.
The resulting broadened spectra are then fit using a Gaussian kernel via the \texttt{pyPIPE3D} fitting code presented in \cite{Lacerda2022pyFIT3D} (as is also used for fitting real MaNGA observations in \cite{Sanchez2022MaNGAPyPIPE3D}). 
In this work, it is noted that there is a lack of high dispersion values seen in the mocks, with maximum velocity dispersion's of 150 km/s observed. 
This is noted as unusual, especially at the higher masses where we would expect the population of slow rotators to dominate. 
While this is expected to be due to an over-representation of late-type systems at higher masses within the TNG100 \citep{Rodriguez-Gomez2019TNGobservations, Huertas-Company2019HubbleTNG}, and TNG50 boxes \citep{Donnari2021QuenchedTNG50}, the true magnitude of this effect may be exaggerated by kinematic measurements that incorporate this systematic effect. 

Within the work of \cite{Nanni2023iMaNGASSP}, galaxies with fewer than 10,000 particles are rejected from their sample, but they do not discuss the minimum number of particles per pixel in their work. 
As a result, it is unclear how this systematic may affect their final result.

It is clear that there are many physical prescriptions that can cause discrepancy in measurements of the LOSVD parameters. 
Never-the-less, it is obvious that at least some of the tension reported within the literature could be alleviated by establishing a convergence threshold in the construction of mock observables. 
This will be imperative for the comparison of simulations as we see new models with new sub-grid models and implementation becoming available. 

\section{Conclusions}
\label{sec:conclusions}

In conclusion, we have presented a number of experiments that demonstrate the importance of constructing mock observations in an unbiased way. 
We have built a series of identically parameterised $N$-body models at a range of spatial sampling resolutions selected to reflect those found within a single hydrodynamical simulation.
Having mock observed these using the \simspin{} code, we show that the recovered observable kinematics will under-predict the magnitude of the line-of-sight velocity dispersion when sampled by fewer than 200 particles. 
This is further demonstrated through a process of bootstrap sampling a high-resolution control model, in order to overcome the assumption that our control models and lower-resolution $N$-body simulations are tracing identical velocity distributions.
We show that the kinematic morphology of a system has an impact on the degree of uncertainty in the recovered LOSVD parameters, particularly in objects dominated by dispersive orbits such as our $N$-body bulge system, while the disc-dominated objects are less strongly affected. 

We implement a new module in v2.7.1 of the \simspin{} code, binning pixels together using a Voronoi tessellation algorithm, as championed by \cite{Cappellari2003Vorbin}, to meet some minimum number of particles-per-pixel prior to measuring the LOSVD with a Gauss-Hermite function. 
Using the $N$-body experiment, we demonstrate that applying the binning algorithm to all kinematic morphologies does not have an impact on observed $\lambda_R$ measurements for disc-dominated objects, but will act to bring those higher dispersion, bulge-dominated systems closer to their true expected value. 
By comparing the spin-ellipticity plane of the EAGLE \texttt{RefL0050N0752} model at increasing numbers of particles-per-pixel, we show that the presence of the slow rotator population stabilises above 200 particles per Voronoi bin, but appears to be missing entirely in unbinned observations. 
Hence, when comparing the stellar observable kinematics within a single model, it is important that this systematic is controlled for.
Considering the impact of this effect on mock surveys of simulations, we can see that the slow rotator fractions of galaxies will likely be under-estimated as a result, 
which will have implications for how new models of feedback are implemented in the next generation of galaxy formation simulations.

These results provide a guideline for the minimum numbers of particles per pixel required to build comparable samples of mock observations.
In Table \ref{tab:uncertainty} we quantify the number of particles required to meet some level of uncertainty for each moment of the Gauss-Hermite and Gaussian LOSVD.
We further demonstrate that this effect is most important for the accurate recovery of the line-of-sight velocity dispersion, $\sigma_{\text{los}}$ in both functional forms.
We conclude that, to reduce the uncertainty a minimum of 200 particles-per-pixel should be used to ensure that mock observations are both comparable to other galaxy models extracted from the same simulation.
Obviously, this extends to comparisons made between different cosmological boxes, as well as the observational samples the mocks are designed to mimic. 
This will be especially prudent for the fair comparison of observation and theory, as we see an increase in popularity of mock surveys, such as iMaNGA \citep{Nanni2023iMaNGASSP} and MaNGIA \citep{Sarmiento2023MaNGIAanalysis}, releasing thousands of mock observations for consistent analysis.

\section*{Acknowledgements}

KH would like to thank the computational theory group at ICRAR-UWA for many thought provoking discussions on this work and the anonymous reviewer for their kind and insightful comments. The clarity of this work has been greatly improved as a result. 
KH acknowledges funding from the Australian Research Council (ARC) Discovery Project DP210101945.
CL has received funding from the ARC Centre of Excellence for All Sky Astrophysics in 3 Dimensions (ASTRO 3D) through project number CE170100013.
LCK acknowledges support by the DFG project nr. 516355818.
Parts of this research were enabled by the Australian Research Council Centre of Excellence for All Sky Astrophysics in 3 Dimensions (ASTRO 3D), through project number CE170100013.
This work was supported by resources provided by the Pawsey Supercomputing Research Centre’s Setonix Supercomputer (\url{https://doi.org/10.48569/18sb-8s43}), with funding from the Australian Government and the Government of Western Australia. 
We acknowledge the Virgo Consortium for making their simulation data available. 
The EAGLE simulations were performed using the DiRAC-2 facility at Durham, managed by the ICC, and the PRACE facility, Curie, based in France at TGCC, CEA, Bruy\'{e}res-le-Ch\^{a}tel.

\section*{Data Availability}
All data used in this paper will be made available through the following GitHub repository, \url{https://github.com/kateharborne/particles_to_pixels_paper}. The $N$-body simulations are provided as the parameter files for Gadget2 and GalIC codes, but the raw particle outputs are also provided in HDF5 format. Mock observations have been constructed using the public \simspin{} code using version 2.9.0 (\url{https://github.com/kateharborne/SimSpin/releases/tag/v2.9.0}). All scripts for building these mocks and the associated figures are included in the paper repository above.  

The \eagle{} simulations and their data are also publicly available through the Virgo database (\url{https://eagle.strw.leidenuniv.nl/wordpress/}). Access to the high-resolution dark matter re-run of \eagle{} \texttt{Ref0050L0752} can be organised upon request.



\bibliographystyle{mnras}
\bibliography{particles_to_pixels} 



\appendix

\section{Quantifying uncertainty}
\label{app:uncertainty}

In \S \ref{sec:convergence}, we consider a bootstrap sample of kinematic observations from two different models at the extremes of the distribution: one structure that is fully supported by rotation (i.e. a thin disc), and another fully supported by dispersion (i.e. a Hernquist bulge). 
We demonstrate that the level of uncertainty on the measure of each mode of the LOSVD decreases with increasing numbers of particles per pixel and use this experiment to advice how many particles may be necessary to recover $\sim$ 95\% of the observed kinematics within some level of acceptable error. 
These values are presented in Table \ref{tab:uncertainty}.

\begin{figure}
    \centering
    \includegraphics[width=\linewidth]{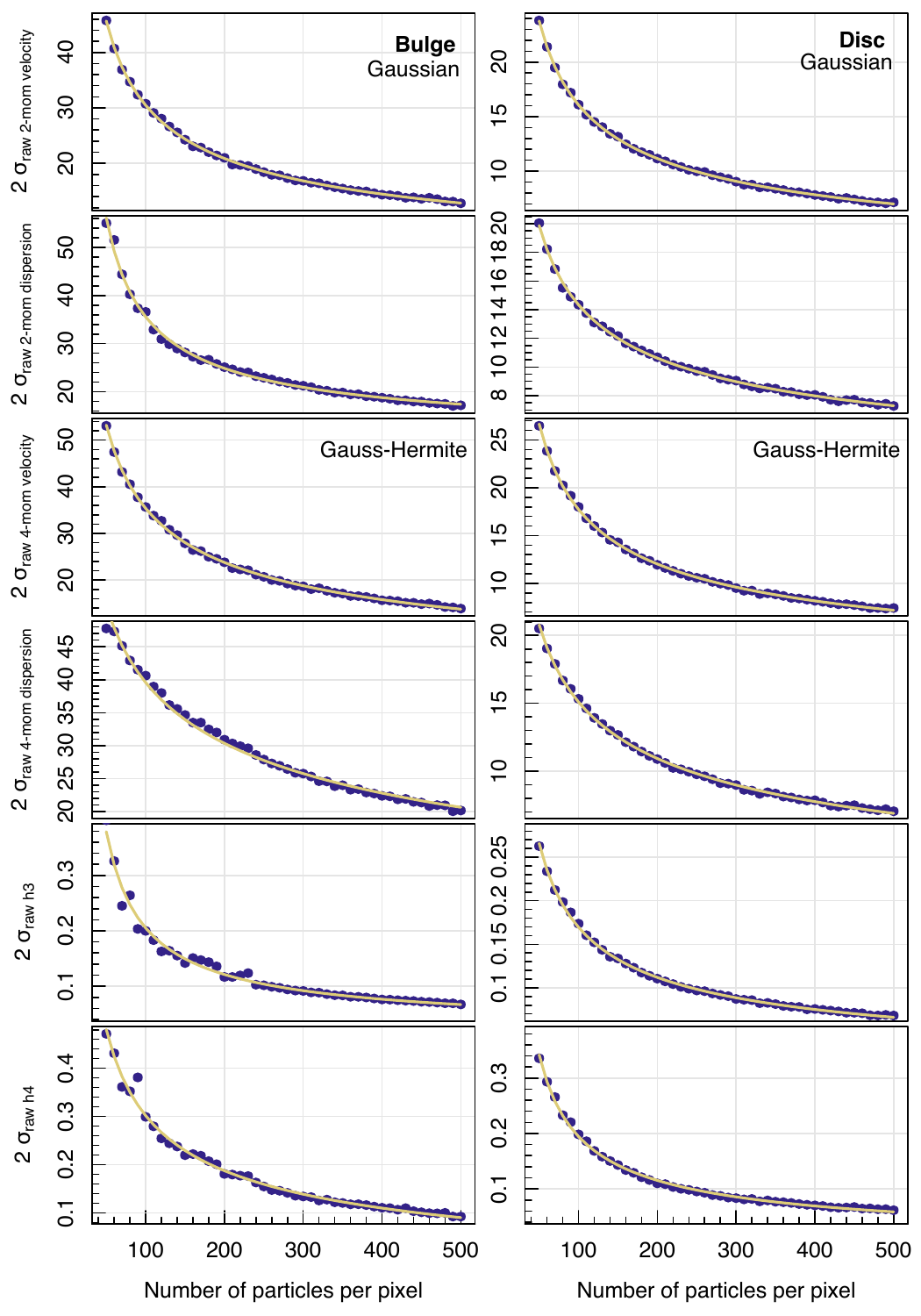}
    \caption{Considering the $2\sigma$ spread of raw differences between the measured bootstrap sampled kinematics and the true value given in the same pixel of the \texttt{control} model, as from Figures \ref{fig:bootstrap_bulge} and \ref{fig:bootstrap_disk}. (Left column) The residual spread for the dispersion dominated model. (Right column) The same, but for the rotation dominated model. Each row shows the distribution for a different mode of the LOSVD: $V$ and $\sigma$ from the Gaussian 2-moment fit; $V$, $\sigma$, $h_3$ and $h_4$ from the Gauss-Hermite 4-moment fit respectively from top to bottom. The yellow lines show the best fit power law to the measured $2\sigma$ distributions shown in purple points.}
    \label{fig:measured_uncertainty}
\end{figure}

To give these minimum number of particle-per-pixel values in this table, we have considered the $2\sigma$-spread of the distribution about the mean for all 8,000 draws made at each number of particles drawn from 50 to 500, in bins of 10.
For clarity, we express the values as a raw difference between the expected value in the high-resolution \texttt{control} pixel and the sampled bootstrap pixel.
The $2\sigma$-spread is measured from this distribution of raw differences. 
We then fit the relationship between the number of particles sampled and the raw $2\sigma$-uncertainty, as shown in Figure \ref{fig:measured_uncertainty}.
A power law of the form below is used for this fit and we use the \texttt{BFGS} algorithm of \citeauthor{BFGS19701}, \citeauthor{BFGS19072}, \citeauthor{BFGS19703} and \citeauthor{BFGS19704} \citeyear{BFGS19701} to optimise the best fitting parameters. 

\begin{equation}
\label{eq:best_fit}
    2\sigma_{\text{kin}} = a \; \text{log}_{10}(N_{p})^{b} + c, 
\end{equation}
where the uncertainty on a kinematic parameter, $2\sigma_{\text{kin}}$, is some power law distribution related to the number of particles per pixel, $N_p$, by some parameters $a$, $b$ and $c$.
The best fitting parameters for each kinematic mode and for each of the extreme model flavours is given in Table \ref{tab:best_fit}.

\begin{table}
\centering
\caption{The best-fit parameters for the measured $2\sigma$-uncertainties for each kinematic parameter and for the bulge and disc idealised systems. These parameters have been optimised using the \texttt{BFGS} algorithm and can be used alongside Equation \ref{eq:best_fit} to give the required number of particles for a given uncertainty in the kinematic mode measured.}
\label{tab:best_fit}
\begin{tabular}{@{}l|c|ccc@{}}
 & \textbf{\begin{tabular}[c]{@{}l@{}}Kinematic \\ mode\end{tabular}} & \textit{\textbf{a}} & \textit{\textbf{b}} & \textit{\textbf{c}} \\ \cmidrule(l){2-5} 
\multirow{4}{*}{\textbf{\begin{tabular}[c]{@{}l@{}}Dispersion\\ dominated\\ bulge\end{tabular}}}  & $V_{\text{2-mom}}$ & 152 & -2.0 & -8.1 \\
 & $\sigma_{\text{2-mom}}$ & 283 & -3.3 & -6.6 \\
 & $V_{\text{4-mom}}$ & 183 & -1.8 & -15.4 \\
 & $\sigma_{\text{4-mom}}$ & 188 & -0.5 & -91.8 \\
 & $h_3$ & 2.15 & -3.39 & -0.01 \\
 & $h_4$ & 1.77 & -1.95 & -0.16 \\ \midrule
\multirow{4}{*}{\textbf{\begin{tabular}[c]{@{}l@{}}Rotation\\ dominated\\ disc\end{tabular}}}    & $V_{\text{2-mom}}$ & 79.7 & -2.0 & -4.3 \\
 & $\sigma_{\text{2-mom}}$ & 56.2 & -1.6 & -4.0 \\
 & $V_{\text{4-mom}}$ & 92.3 & -2.0 & -6.0 \\
 & $\sigma_{\text{4-mom}}$ & 62.8 & -1.1 & -15.0 \\
 & $h_3$ & 1.24 & -2.75 & -0.01 \\
 & $h_4$ & 2.07 & -3.249 & -0.02
\end{tabular}
\end{table}

\section{Other resolution considerations}
\label{app:highres-dm}

In Section \ref{sec:convergence}, we consider the change in kinematic parameter space as mock observations are gridded to reach higher numbers of particles-per-pixel for the \eagle{} \texttt{RefL0050N0752} volume. 
This simulation is an intermediate resolution run in which the dark matter particles are 5.389 times more massive than the gas particles from which stars are born. 
While this reduces the computational complexity by reducing the number of calculations required for the dark matter component, the reduced number of dark matter particles at fixed halo mass leads to spurious kineomorphological evolution.
As noted in the works of \cite{Ludlow2019EnergySizes, Ludlow2021Spuriousparticles} and \cite{Wilkinson2023impactdiscs}, this will subtly vary standard scaling relations, increasing the sizes of galaxies and reducing their stellar rotational velocities. 
This leads us to ask whether the effects observed in our work are simply due to these resolution effects.  

\begin{figure*}
    \centering
    \includegraphics[width=\linewidth]{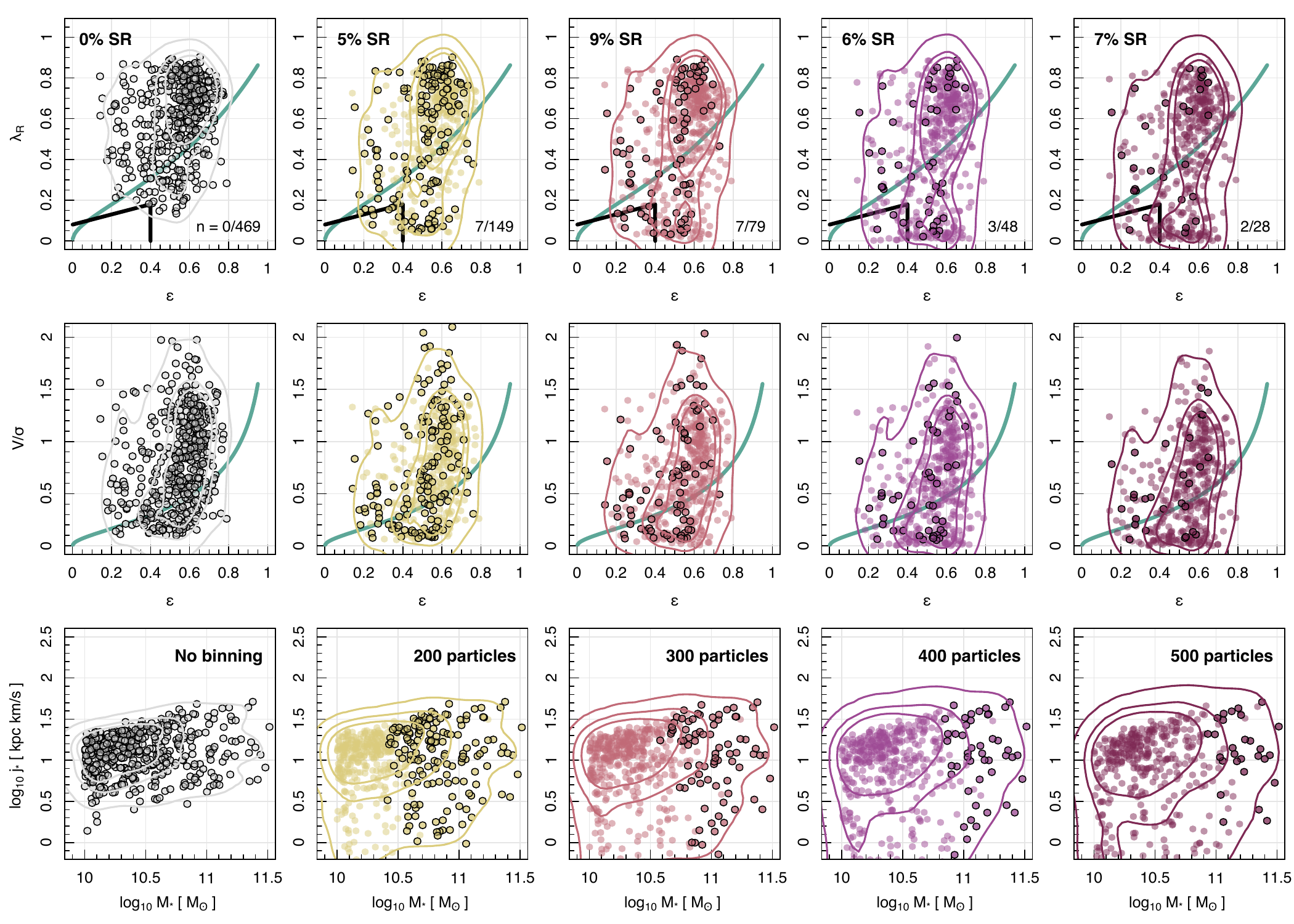}
    \caption{For galaxies extracted from the HiResDM \eagle{} simulation, we consider the same extracted kinematic properties as in Fig \ref{fig:eagle_kinematics}. (Top row) The spin-ellipticity ($\lambda_R$-$\varepsilon$) plane, where the black line shows the slow/fast rotator boundary as suggested by \protect\cite{Cappellari2016AnnualReview} and the green line shows the theoretical position of an edge-on oblate system with given $\varepsilon$ and anisotropy described by $\beta$ = $0.7$ $\times$ $\varepsilon$; (middle row) the V$/\sigma$-$\varepsilon$ distribution with the same definition magenta line as on the top row; and (bottom row) the projected stellar angular momentum $j_*$ as a function of stellar mass. Points are individual observations of galaxies extracted from \eagle{} and the overlaid contours show the distribution of 95\%, 68\% and 50\% of all the data. Points outlined in black have at least 50 pixels within the measurement radius. We give the number of these black outlined galaxies in the lower right of the top column (where $n$ = number classes SR / total number). The left column in grey points show measurements made without any restrictions on the number of particles-per-pixel from which the kinematics have been measured, i.e. unbinned. The following columns show the same kinematics measured from mock images that have been Voronoi binned to meet some target number of particles-per-pixel, as listed in the bottom right hand corner of each row of images. As can be seen from the slow rotator fraction listed in the upper left of each row, the effect of binning versus not binning has an important affect on the distributions, but once at 200 particles per pixel and above, there is only small amounts of variation between the panels.}
    \label{fig:eagle_kinematics_hires}
\end{figure*}

To answer this, we took the \eagle{} simulation run by \cite{Ludlow2021Spuriousparticles}, in which the number of dark matter particles sampling the underlying potential was increased by a factor of 7. 
In this simulation, the resolution of the gas particles ($m_g$ = $1.8 \times 10^{6}$ \msol) and volume of the box (50 cMpc) is consistent with the \texttt{RefL0050N0752} volume, but the mass of the dark matter particle is reduced to $1.39 \times 10^{6}$ \msol.
Following the naming convention of \cite{Ludlow2021Spuriousparticles}, this higher resolution volume is called the ``HiResDM''.
Using the same technique as in \S \ref{sec:convergence} for the \texttt{RefL0050N0752} volume, we extracted 469 galaxies from the HiResDM simulation and make mock observations of the extracted galaxies from the same fixed inclination of 70 degrees, fitting each LOSVD with a Gaussian kernel as described by Eq. \ref{eq:Gaussian}. 
Mocks are then Voronoi binned as before into pixels containing 200, 300, 400 and 500 particles. 
We then measured the same kinematic properties from these mock observations as for the lower resolution volume. 
In Figure \ref{fig:eagle_kinematics_hires}, we demonstrate these results. 

Although we do see a change in the parameter space between Figures \ref{fig:eagle_kinematics} and \ref{fig:eagle_kinematics_hires}, the general effect of binning mock observations is consistent. 
We see that, when the number of particles per pixel are not controlled for, in the unbinned case, the affects of shot noise in our measurements cause observed kinematic distributions to look different.
Galaxies sit higher in each space, as dispersions are underestimated.
Once observations have been voronoi binned, we see the distribution largely stabilise. 
With regards to the $f_{SR}$, the value sits at $\sim 5-9\%$, slightly higher than our "LowResDM" \eagle{} experiment shown in Figure \ref{fig:eagle_kinematics}.
This is consistent with the results in the main body of this paper, however, demonstrating that binning of particles in pixelated mock observations is important for reliable comparison between simulations and observations.
These effects are independent of the underlying resolution of the simulation in question. 


\bsp	
\label{lastpage}
\end{document}